\documentclass[landscape,usenatbib,usegraphicx,twocolumn]{mn2e}
\usepackage{graphicx}
\usepackage{colordvi}
\renewcommand{\thefootnote}{\fnsymbol{footnote}}

\def\ba{\begin{eqnarray}}
\def\ea{\end{eqnarray}}
\newcommand{\be}{\begin{eqnarray}}
\newcommand{\ee}{\end{eqnarray}}

\def\theequation{\thesection.\arabic{equation}}
\title[Cosmological parameter estimation in presence of extragalactic point sources]
{A Multifrequency approach of the cosmological parameter estimation in presence of extragalactic point sources}

\author[D. Paoletti, N. Aghanim, M. Douspis, F. Finelli, G. De Zotti, G. Lagache and Aur\'elie P\'enin]
{\parbox{\textwidth}{D.~Paoletti \thanks{paoletti@iasfbo.inaf.it}$^{1,2}$, 
N.~Aghanim \thanks{nabila.aghanim@ias.u-psud.fr}$^{3}$,
M.~Douspis \thanks{marian.douspis@ias.u-psud.fr}$^{3}$, 
F.~Finelli \thanks{finelli@iasfbo.inaf.it}$^{1,2}$,
G.~De Zotti \thanks{gianfranco.dezotti@oapd.inaf.it}$^{4,5}$,
G.~Lagache \thanks{guilaine.lagache@ias.u-psud.fr}$^{3}$ and
A.~P\'enin \thanks{aurelie.penin@ias.u-psud.fr}$^{3}$}\vspace{0.4cm}\\
\parbox{\textwidth}{$^1$ INAF-IASF Bologna, Istituto di Astrofisica Spaziale e Fisica Cosmica 
di Bologna \\
Istituto Nazionale di Astrofisica, via Gobetti 101, I-40129 Bologna - Italy \\
$^2$ INFN, Sezione di Bologna,
Via Irnerio 46, I-40126 Bologna, Italy \\
$^3$ IAS Institut d'Astrophysique Spatiale\\
B\^at. 121 Universite Paris Sud 11 \& CNRS, 91405 Orsay Cedex, France \\
$^4$ INAF, Osservatorio Astronomico di Padova,
Vicolo Osservatorio 5, I-35122 Padova, Italy \\
$^5$ SISSA,
Via Bonomea 265, I-34136 Trieste, Italy \\
}}

\begin{document}

\def\lsim{\,\lower2truept\hbox{${< \atop\hbox{\raise4truept\hbox{$\sim$}}}$}\,}
\def\gsim{\,\lower2truept\hbox{${> \atop\hbox{\raise4truept\hbox{$\sim$}}}$}\,}

\maketitle

\begin{abstract}
We present a multifrequency approach which optimizes the constraints on cosmological 
parameters with respect to extragalactic sources and secondary anisotropies contamination on small scales.
We model with a minimal number of parameters the expected dominant contaminations in  intensity, such as 
unresolved point sources and the thermal Sunyaev-Zeldovich effect. 
The model for unresolved point sources, either Poisson distributed or clustered, uses data from {\it Planck} early
 results.
The overall amplitude of these contributions
are included in a Markov Chain Monte Carlo analysis for the estimate of cosmological parameters.
We show that our method is robust: 
as long as the main contaminants are taken into account the constraints on the cosmological
parameters are unbiased regardless of the realistic uncertainties on the contaminants. 
We show also that the two parameters modelling unresolved points sources are not prior dominated.
\end{abstract}


\raggedbottom
\setcounter{page}{1}

\section{Introduction}

The CMB anisotropy data constitute a fundamental tool to test the standard
cosmological model and its extensions.
In particular small scale CMB anisotropies have a great importance 
in cosmology and are one of the current frontiers in
CMB observations. 
Many experiments have been dedicated to the observations of small scale
anisotropies. The {\it Planck} satellite \citep{Bluebook} will cover all the scales up to $\ell\sim 2500$ but there
are also ground based experiments observing smaller regions of the sky but with higher angular resolutions, such as: 
ACBAR \citep{ACBAR}, CBI \citep{CBI}, QUaD \citep{QUAD}, SPT \citep{SPT}, ACT \citep{ACT}.
At these scales, the Silk damping suppresses the primordial CMB contribution with respect to extragalactic 
contamination and secondary anisotropies. These contaminations affect
the constraints on cosmological parameters derived from small scale data introducing biases and therefore
it is necessary to properly account for them in the data analysis. We present
an approach which has been developed to optimize, with respect to the contamination by
point source contributions, the constraints on cosmological 
parameters which could be obtained from high resolution multi-frequency data.\\ 
The problem of extragalactic sources and secondary anisotropies contamination of the angular power spectrum 
and consequentely of the cosmological parameters has already been addressed
in several recent articles \citep{Douspis2006,TaburetBias,TaburetSZ,EfstathiouGratton,Millea,IRCTemplate,SZGeorge}, 
together with the specific tools developed by the data analysis groups of the 
most recent small scale experiments \citep{SPT2011,ACT2010}. \\

The approach we present in this article consists in modelling the astrophysical contributions to 
the angular power spectrum with minimal parametrizations, where 
minimal relates to the number of parameters necessary to describe the signals, and considering 
the frequency dependences of the different contributions.
This is the most important difference with respect to previous approaches
\citep{EfstathiouGratton,Millea,IRCTemplate,SPT2011,ACT2010}. These parametrizations are included 
as additional contributions
to the primary CMB in the Markov Chain MonteCarlo analysis to constrain cosmological parameters.
The parameters characterizing the astrophysical contributions are included in the analysis together with the cosmological
ones. As described in the following, we focus on the frequencies: $70,100,143,217,353$ GHz. 
We use the available data from {\it Planck} early results \citep{PlanckRadio,PlanckCIB}
to derive the parametrizations to describe the astrophysical signals. When no data are available, we 
rely on predictions from theoretical models and empirical simulations.\\

In the following, we use a fiducial cosmological model with the parameters from \cite{WMAP7} which are reported in the third column of  Table 4. The polarization
from point sources and SZ effect is negligible we thus do not take it into account. We also do not account
for residual signal, in intensity nor in polarization, from our Galaxy. \\

The article is organized as follows.  In Section 2 we describe the astrophysical signals
and we derive the parametrizations of their contributions to the angular power spectrum.
In Sect 3 we describe the analysis method and the frequency channel combinations.
In Section 4 we present the results of the Markov Chain MonteCarlo (MCMC) analysis for the cosmological
and astrophysical parameters. In section 5 we present the discussion and conclusions.

\section{Astrophysical contaminations}

One of the major contributions of astrophysical contamination on small scales, and in the regions of the
sky clean from the Galactic emission,
is given by discrete sources and in particular galaxy clusters and
the extra-galactic point sources point sources. 
The former contributes through the Sunyaev Zeldovich (SZ) effect
\citep{TSZOriginal,KSZOriginal} whereas the latter contribution depends on the frequency. 
We have considered only the contribution of the thermal SZ effect since the kinetic SZ effect has a smaller amplitude.
For the point-source contamination we considered different contributions
depending on the frequency. We aim in this section at parametrising the different contributions by
reducing to their minimum the number of parameters necessary to characterize the signals, and by
including the frequency dependence in these parameters.

\subsection{Thermal Sunyaev Zeldovich effect}

The thermal SZ effect (TSZ) is a local spectral distortion of the CMB
given by the interactions of CMB photons with the electrons of the hot
gas in galaxy clusters \citep{TSZOriginal} encountered during the
propagation from the last scattering surface.

The TSZ contribution to the CMB signal strongly depends on the cosmological model and the 
cluster distribution through the dark matter halo mass function. It also
depends on the cluster properties through the
projected gas pressure profile. The TSZ angular power spectrum can be 
computed using these two ingredients (e.g. \cite{KomatsuSeljak,AghanimReview}).
It was shown in \cite{TaburetBias} that the SZ contribution 
to the signal can be analysed in terms of a residual contribution to the total signal
when the detected clusters are masked out from the maps, 
or in the contrary the TSZ signal can be analysed together with the CMB one
without taking out the detected clusters providing the total SZ power spectrum is
properly modelled, see \cite{TaburetSZ}. The latter approach is the one that is used in
the CMB experiments like WMAP \citep{WMAP7}, ACT \citep{ACT,ACT2010} and SPT \citep{SPT,SPT2011}. 
As discussed in \cite{TaburetSZ}, we can
consider three possible parametrizations for the TSZ power
spectrum. However, the authors showed that the parametrization which
represents physically the best the TSZ signal and at the same time does not bias
the cosmological parameters is the one derived in \cite{KomatsuSeljak}. It
considers a template for the spectral shape, $\hat C_\ell$, and a normalisation which depends on
the parameters $\sigma_8$ and $\Omega_b^2 h^2$. According to these results, we chose to use: 
\be
C_\ell^{\mathrm{SZ}}(\nu)=A_{\mathrm{SZ}}g(\nu)\sigma_8^7\Omega_{\mathrm b}^2 h^2 \hat C_\ell,
\label{SZR}
\ee
where $A_{\mathrm{SZ}}$ accounts
for the uncertainty in the normalisation.  The frequency dependence
of the signal for the TSZ is given by the analytical function $g(\nu)$\footnote {In the non-relativistic approximation, it is given by: $g(\nu)=-\Big(x\frac{e^{x}+1}{e^{x}-1}-4\Big)$ with $x=\frac{\nu}{56.78\mathrm{GHz}}$.}. 
In the following, we used a fiducial TSZ power-spectrum shape computed 
using the \cite{Tinker2008} mass function and a pressure profile provided in 
\cite{Arnaud2009}.

\subsection{Point source contribution}
An experiment like {\it Planck} with its moderate angular resolution is not
optimized for point source detection. However, {\it Planck}, thanks to
its full sky coverage and its 9 frequencies, is detecting a large
number of sources with the detection thresholds reported in Table
\ref{tab_PSP} and taken from \cite{PlanckCIB}. 
For the CMB analysis, the contribution of detected
sources is removed by masking them in the maps. However, the
contribution of unresolved point sources remains and needs to be
properly dealt with since it acts like an unavoidable noise that
impacts the angular power spectrum amplitude and the cosmological
information contained in it. \\

In the frequency range of interest for our study the main contributions
from extragalactic point sources are those of the radio galaxies which
dominate the low frequency channels and the infrared ones which
dominate the higher frequencies.  Together with these two main
populations there is a contribution from anomalous objects like
Gigahertz Peaked Spectrum, Advection Dominated Accretion
Flows/Adiabatic Inflow Outflow Solution and radio
afterglows of Gamma Ray Burst
\citep{Toffolatti98,Toffolatti2005,DeZotti2005,DeZotti2005Planck}.\\

Radio galaxies, typically BL LACs and Flat Spectrum Radio Quasars
present a flat spectrum $S_\nu\propto \nu^\alpha$ with
$\alpha > {-0.5}$ \citep{Danese1987}. 
There is also a contribution of steep spectra radio sources
which typically have fluxes $S_\nu\propto \nu^\alpha$ with $\alpha\sim
{-0.7-(-1)}$.  The anomalous objects have only a minor contribution to
the total of radio galaxy power at the arcmin scale (see for example
\cite{Toffolatti2005} for detailed discussion).  Infrared galaxy
emission presents a steep spectrum $S_\nu\propto \nu^{3.5}$ and is due
to the UV light from star formation reprocessed by dust in
galaxies. Infrared galaxies have been lengthly studied and observed in
the recent past especially by CMB-dedicated experiments like ACT (\cite{ACT} and references therein),
SPT \citep{SPT} and {\it Planck} \citep{PlanckCIB} but also by 
Spitzer \citep{SPITZER}, Blast \citep{BLAST}, and Herschel \citep{HERSCHEL}.\\

In our range of frequencies and for our considered thresholds, the radio source contribution 
to the angular power spectrum can be described only by a Poissonian term. 
The situation is quite different for infrared galaxies. The latter have in fact a strong 
clustering contribution to the angular power specturm  and a careful modeling and
parametrisation of their clustering term are necessary \citep{PlanckCIB,Penin}. We thus consider the following
contributions from point sources: 
\ba 
70\,\mathrm{GHz} &\rightarrow & \mathrm{Poisson(Radio)}\nonumber\\ 
100\,\mathrm{GHz} &\rightarrow & \mathrm{Poisson(Radio)+IR\,Clustering}\nonumber\\ 
143\,\mathrm{GHz} &\rightarrow & \mathrm{Poisson(Radio+IR) +IR\,Clustering}\nonumber\\ 
217\,\mathrm{GHz} &\rightarrow & \mathrm{Poisson(Radio+IR)+IR\,Clustering}\nonumber\\ 
353\,\mathrm{GHz} &\rightarrow &
\mathrm{Poisson(Radio+IR)+IR\,Clustering}\nonumber\,.  
\ea

In the following, we describe the different contributions from point sources to the CMB anisotropy, 
we detail the proposed parametrisations and then provide the associated angular power
spectrum.

\subsection{Poissonian contribution}
The Poisson contribution of point sources is given by their random
distribution in the sky.
The coefficients of spherical harmonics of a Poissonian distribution
of a population of sources in the sky with flux density $S$ are
\citep{TegmarkEfstathiou}:
\begin{displaymath}
 \langle a_{\ell m}\rangle =
\left\{\begin{array}{ll}
 \sqrt{4\pi}\bar{n}S\quad \mathrm{for}\quad\ell=0\\
0\quad \mathrm{for}\quad\ell\ne 0,\forall\, m
\end{array}  \right.
\end{displaymath}
where $\bar{n}=N/4\pi$ is the mean number of sources per steradian
with flux density $S$.  The angular power spectrum is given by
$C_\ell=\langle|a_{\ell m}|^2\rangle-|\langle a_{\ell
  m}\rangle|^2=\bar{n}S^2\,,$ which can be generalized to
$C_\ell=\Sigma_i\bar{n_i}S_i^2$ if we consider sources with different
flux densities.  Extrapolating the continuum limit in flux up to the detection
threshold, $S_{\mathrm{max}}$, the angular power spectrum for a Poissonian distribution of
sources in sky is given by a flat $C_\ell$: 
\begin{equation}
C_\ell^{\mathrm{Pois}}=\int_0^{S_{\mathrm{max}}} \frac{\mathrm{d}N(S)}{\mathrm{d}S} S^2 \mathrm{d}S \,,
\label{ClPSP}
\end{equation} 
where $\frac{{\mathrm d}N(S)}{{\mathrm d}S}$ are the source number counts.
The flat spectral shape of the Poissonian contribution makes it easy to
parametrize: It requires a simple amplitude term. One possible choice of parametrization is to use a different free
parameter for the amplitude for each frequency as in \cite{EfstathiouGratton}. This would require in our case for the radio sources alone 5 parameters for the Poisson term.  Since we aim at reducing the number of parameters to the minimum, we construct a single amplitude parameter for the total Poisson contribution and accounting for the frequency dependence. This frequency dependence is complicated by the dependence on the detection
threshold $S_{\mathrm{max}}$ and
thus no analytical expression can be constructed. We therefore reconstruct from the data
and from the galaxy evolution models the necessary information to derive the single parameter for the
Poisson term.

\begin{figure}
\begin{tabular}c
\includegraphics[scale=0.76]{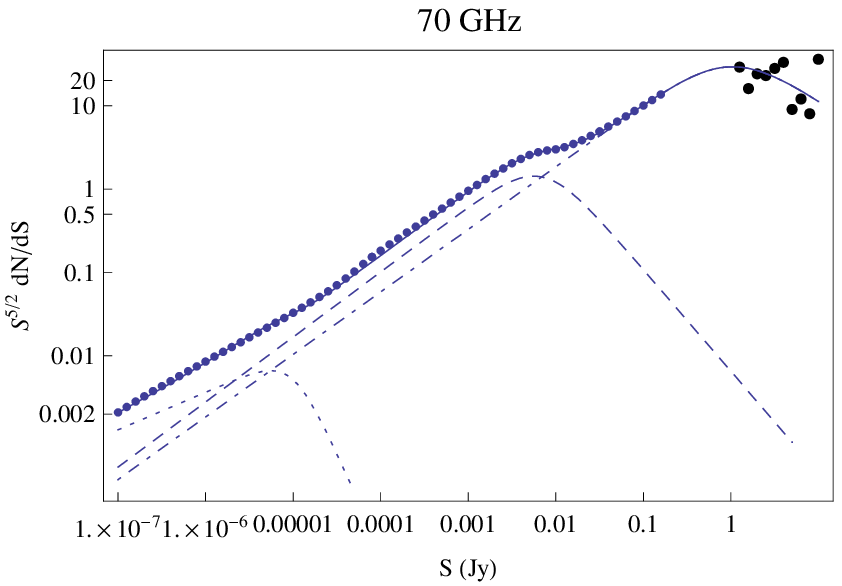}\\
\includegraphics[scale=0.76]{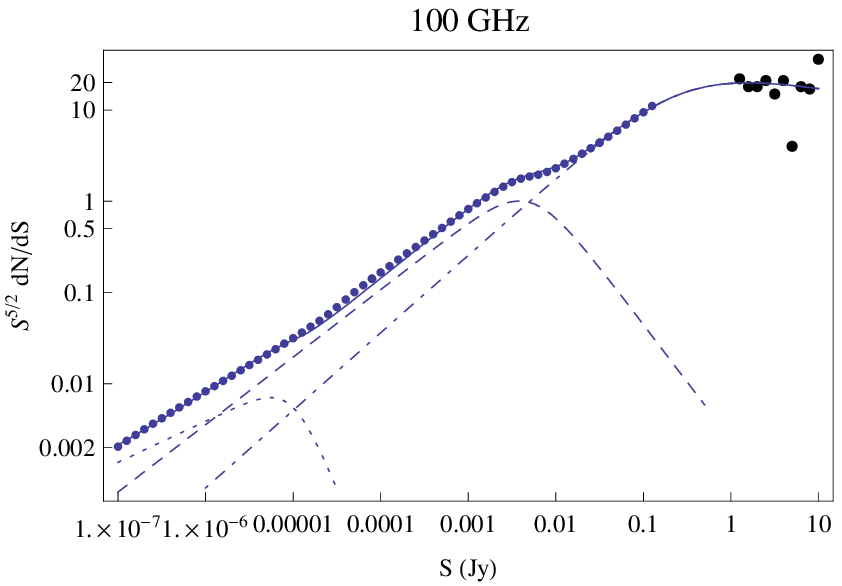}\\
\includegraphics[scale=0.76]{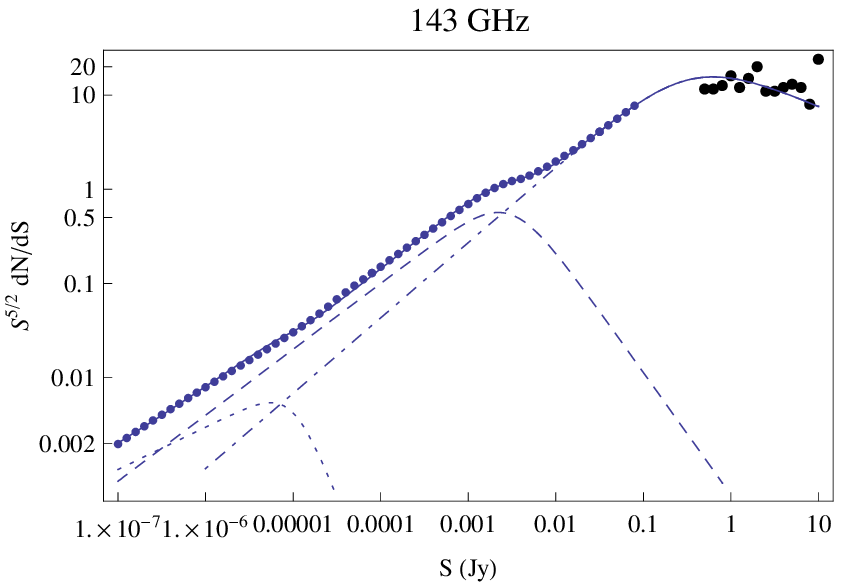}\\
\end{tabular}
\caption{Fits of the hybrid model/data number counts for each frequency. 
The solid line is the empirical fit whereas the dots represent the 
De Zotti et al. model number counts (thin light dots) and Planck data (dark thick dots).
The three curves under the fit are the three contributions (Low, Med and High, respectively dotted, dashed and dot-dahsed lines)
 which sum in the total fit. From top to bottom: $70$ GHz, $100$ GHz, $143$ GHz.}
\label{PSFIT}
\end{figure}

\begin{figure}
\begin{tabular}c
\includegraphics[scale=0.76]{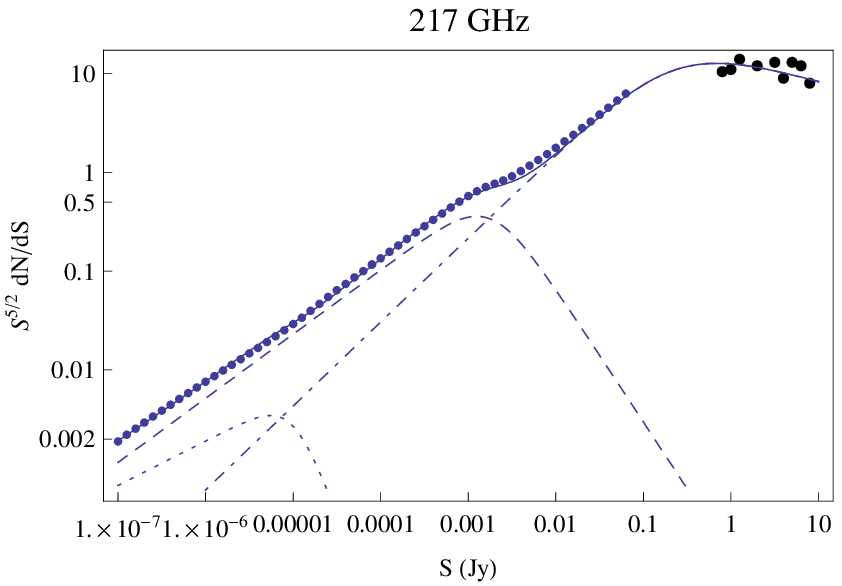}\\
\includegraphics[scale=0.76]{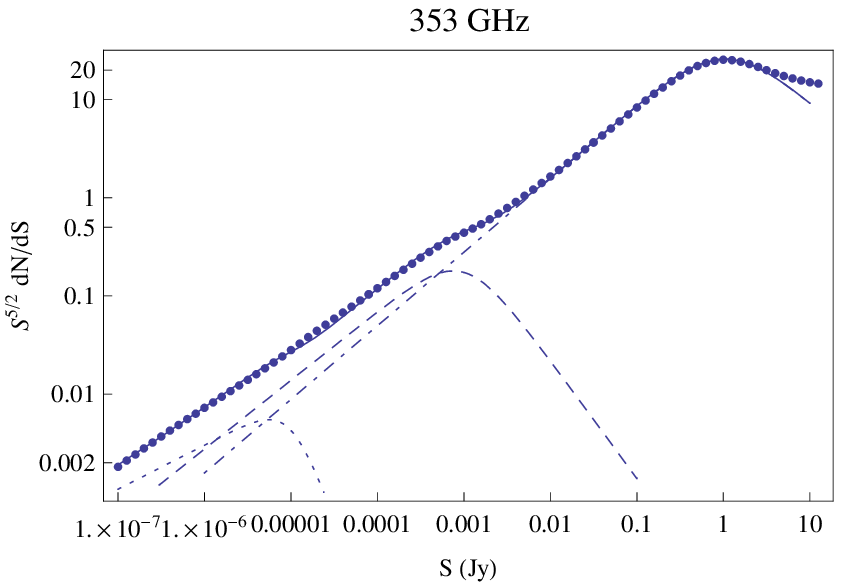}\\
\includegraphics[scale=0.76]{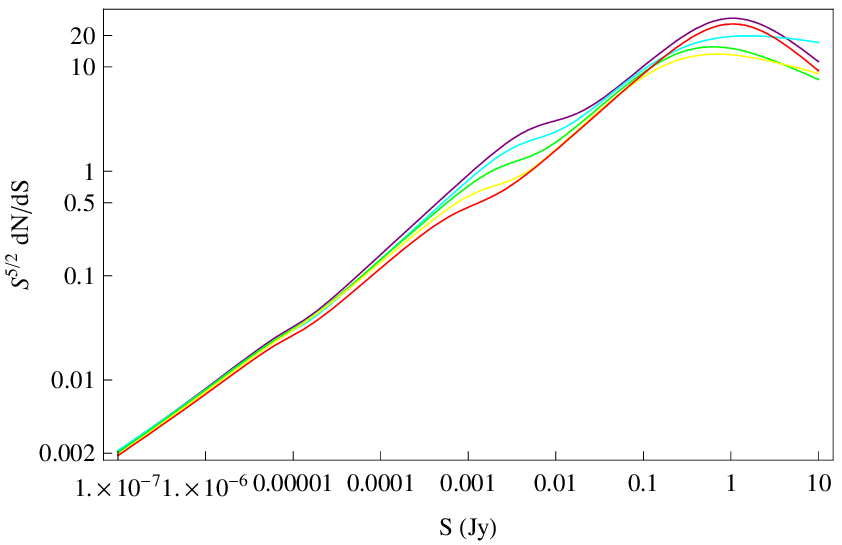}\\
\end{tabular}
\caption{Fits of the hybrid model/data number counts for each frequency. 
The solid line is the empirical fit whereas the dots represent the 
De Zotti et al. model number counts (thin light dots) and Planck data (dark thick dots).
The three curves under the fit are the three contributions (Low, Med and High, respectively dotted, dashed and dot-dahsed lines)
 which sum in the total fit. From top to bottom: $217$ GHz, $353$
  GHz. Bottom  is the comparison of the fits for the
  different frequencies: $70$ GHz is purple, $100$ GHz is cyan, $143$
  GHz is green, $217$ GHz is yellow, $353$ GHz is orange.}
\label{PSFIT2}
\end{figure}

Both radio and infrared galaxies contribute with a Poissonian term. We treated the former 
by fitting the number counts and then integrating on the flux density whereas for the latter we used
the available results from \citep{Bethermin2010, PlanckCIB}. The two contributions are then 
summed in a single Poissonian term.
The expected Poissonian contribution for each frequency can be
computed using the number counts given a detection threshold. We used
the predicted number counts for radiogalaxies from the model of
\citep{DeZotti2005,DeZotti2005Planck}. It is a good representation of
low flux part of the observed number counts but the comparison with the {\it
  Planck} Early Release Compact Source Catalogue (ERCSC) showed a bias
at high flux densities with the De Zotti et al. model
overpredicting the counts \citep{PlanckRadio,Tucci}.  Since the ERCSC number counts are limited to bright flux densities, 
we constructed composite number counts combining the De Zotti
model predicted counts, for lower flux densities roughly below $0.2-0.3$ Jy (thin dots in Figs. \ref{PSFIT} and  \ref{PSFIT2}),
whereas for flux densities greater than $1$ Jy we used the actual {\it Planck} measurements
reported in \cite{PlanckRadio} (thick dots in Figs. \ref{PSFIT} and  \ref{PSFIT2}). 
 
At 353 GHz, the radio sources 
are subdominant with respect to infrared sources. In order to compute the associated Poisson term for the 353 GHz channel 
we therefore used the predictions of the De Zotti et al. model that we rescaled to account for its likely
overprediction, similar to that at 143 and 217 GHz. In practice, we have extrapolated the correction 
factors for the radiogalaxy Poissonian contribution to the angular power spectrum of the 143 and 217 GHz channels, $0.93$ and $0.92$
respectively, and found a correction factor of $0.91$ for the Poissonian angular power spectrum of the 353 GHz channel.

The obtained number counts\footnote{The differential
  number counts are normalized to the Euclidean ones ($\frac{{\mathrm
      d}N(S)}{{\mathrm d}S}\propto S^{-2.5}$).}
are represented for each frequency by the weighted sum of three different contributions:
one at low $F_{\nu}^{\mathrm{Low}}$,
one at intermediate $F_{\nu}^{\mathrm{Med}}$, and one at high $F_{\nu}^{\mathrm{High}}$
flux densities. The number counts are given by: 
\be
S^{-2.5}\frac{\mathrm{d}N(S)}{\mathrm{d}S}^{\mathrm{tot}}_{\nu}=\Sigma_{i=\mathrm{Low,Med,High}}w_{\nu}^i\,F_{\nu}^i 
\ee
Each term $F_{\nu}^{\mathrm{Low, Med, High}}=S^{-2.5}\frac{{\mathrm
 d}N(S)}{{\mathrm{d}}S}$ writes as $F_{\nu}^{\mathrm Low, Med,
High}=\frac{A\,S^\alpha}{(C+B\,S^\beta)}$. 
For each term and each frequency, the exponents $\alpha$ and $\beta$
and the coefficients $A$, $B$ and $C$, together with the weights of the sum $w_{\nu}^{\mathrm
  Low, Med, High}$ are all adjusted by comparing them to the
composite number counts. Table 5 summarises the set of parameters used
to construct the composite source counts and the fit to the hybrid model/data counts is shown as the
thin line in Figs. \ref{PSFIT} and  \ref{PSFIT2}. 
We note the very good agreement between the analytical fit and the theoretical/measured points. \\

\begin{figure}
\begin{tabular}c
\includegraphics[scale=0.76]{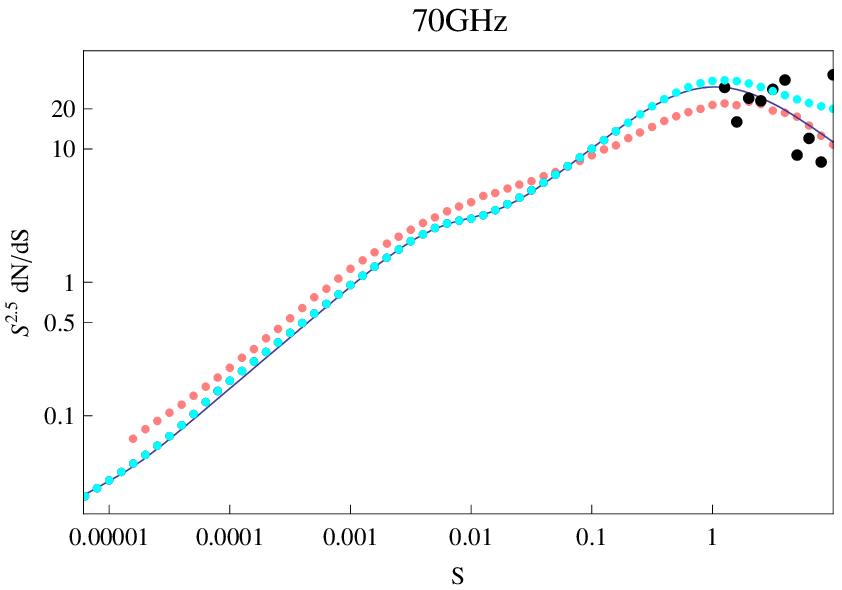}\\
\includegraphics[scale=0.76]{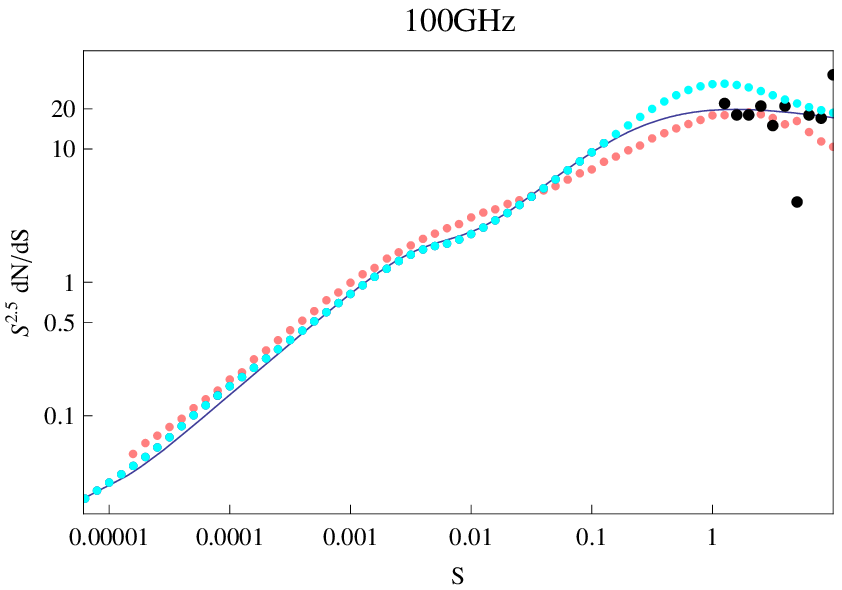}\\
\includegraphics[scale=0.76]{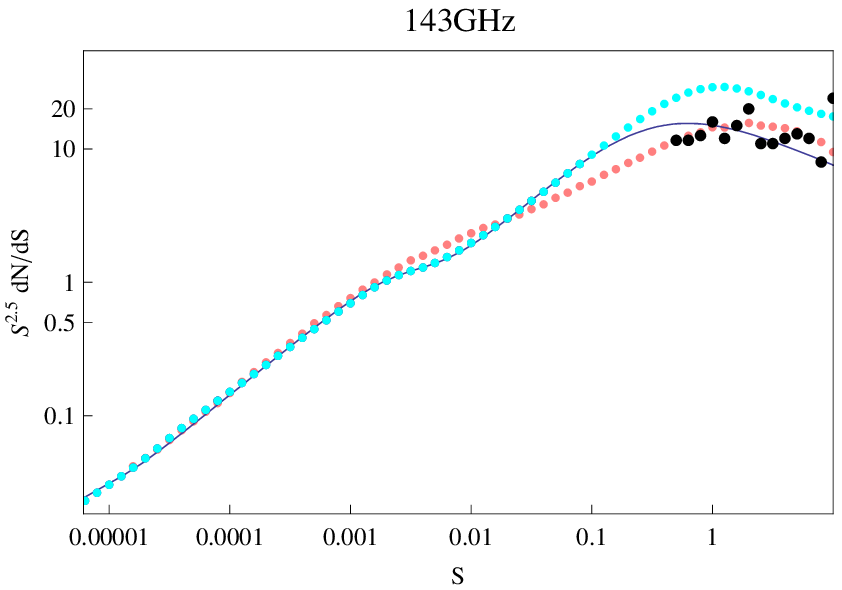}\\
\end{tabular}
\caption{Comparison of the fits of the hybrid model/data number counts with the Tucci et al. model
for each frequency. 
The solid line is the empirical fit whereas the cyan dots represent the 
De Zotti et al. model number counts (thin light dots) and Planck data (dark thick dots).
The Tucci et al. model is represented by pink dots. From top to bottom: $70$ GHz, $100$ GHz, $143$ GHz.}
\label{TUCCI1}
\end{figure}
\begin{figure}
\begin{tabular}c
\includegraphics[scale=0.76]{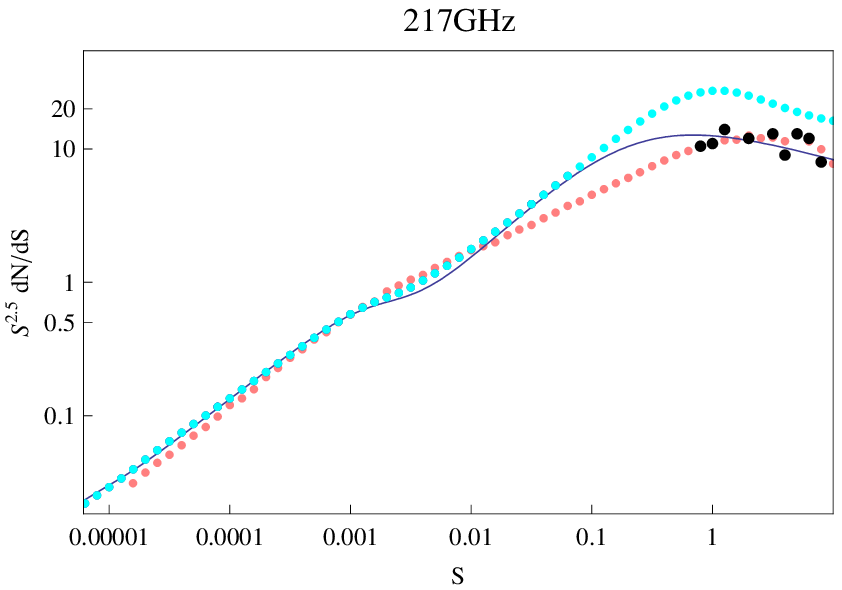}\\
\includegraphics[scale=0.76]{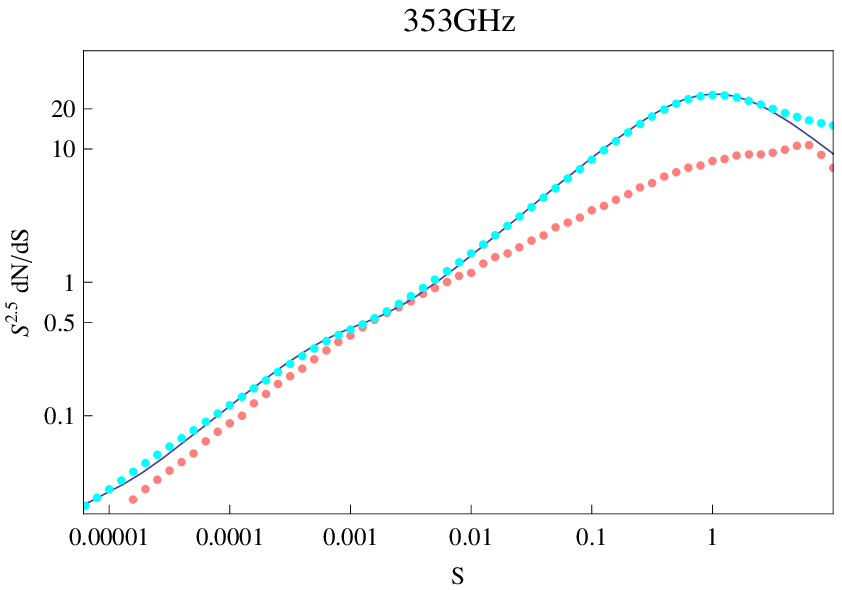}\\
\end{tabular}
\caption{Comparison of the fits of the  hybrid model/data number counts with the Tucci et al. model
for each frequency. 
The solid line is the empirical fit whereas the cyan dots represent the 
De Zotti et al. model number counts (thin light dots) and Planck data (dark thick dots).
The Tucci et al. model is represented by pink dots. From top to bottom: $217$ GHz, $353$ GHz.}
\label{TUCCI2}
\end{figure}

In Figs. \ref{TUCCI1} and \ref{TUCCI2} we compare our hybrid number counts to those of \cite{Tucci} based also on the ERCSC. Our fit
overpredicts the number counts in the range $0.01-1$ Jy with respect to the Tucci et al. model. 
This overprediction is due to the fact that the flux density at we match the predictions and the data is 0.01 Jy whereas Tucci et al. chose a lower value.

Using the obtained composite counts we compute the Poissonnian power
spectra at the considered frequencies. They are given by following
expressions:
 
\ba
C_{\nu}^{\mathrm RadioPois} &=&\Sigma_{i=1,3} P_{i,\,\nu}S_{\mathrm{max},\,\nu}^{\chi_{i,\,\nu}} \nonumber\\
&& \times 2F_1[a_{i,\,\nu},b_{i,\,\nu},c_{i,\,\nu},d_{i,\,\nu}  S_{\mathrm{{max},\,\nu}}^{e_{i\,\nu}}(\nu)]
\label{RadioP}
\ea

where $2F1$ are the Hypergeometric functions of second type \citep{AbramowitzStegun}, the sum on $i$ is over 
the three contributions Low, Med and High, and the coefficients and 
exponents depend on the frequency and are reported in Table 6.
The values of the Poissonian contribution to the angular power spectrum that 
we obtained from our fitted number counts  are reported in the third column of Table \ref{tab_PSP}. For comparison in the fourth
column we report the values computed from the original De Zotti-model fits. As
expected from the number counts comparison,
we have a higher Poissonian contribution than the one obtained with the Tucci et al. model, namely 30\% higher at 70 and 100 GHz, 40 and 60\% at 143 and 217 GHz respectively. At 353 GHz we predict an amplitude twice larger.

Together with the term coming from the radio sources it is necessary to
consider the Poissonian contribution of infrared galaxies which is relevant only for the higher frequency channels
 the 143, 217 and 353 GHZ. The values of the
Poissonian infrared contributions at these frequencies are computed from the \cite{Bethermin2010} model
 in \cite{PlanckCIB} and reported  in the fifth column of Table \ref{tab_PSP}. 
We summed the two Poisson terms for radio and infrared in the frequencies in a single Poissonian term:
\be
C_{l,\nu}^\mathrm{{PS,TOT}}=A_\mathrm{{PS}} C_{l,\nu}^\mathrm{{PS,Radio+IR}}\,,
\ee
where $C_{l,\nu}^\mathrm{{PS}}$ is the Poissonian contribution at the different frequencies whereas $A_\mathrm{{PS}}$ which incorporates deviations of real data from this model remains the same for all the frequency considered.

\begin{table}
\centering
\small\addtolength{\tabcolsep}{-4pt}
\begin{tabular}{|c|c|c|c|c|}
\hline
$\rm{Channels}$& $S_{\rm max}$ & $C_{\ell,{\rm Hybrid}}^{\rm Pois,Radio}$  & $C_{\ell\,{\rm Original}}^{\rm Pois,Radio}$  &$C_{\ell\,\mathrm{IR}}^{\rm Pois}$ \\ 
\hline
& $\rm(J_y)$ & $\rm{(\mu K^2)}$  & $\rm{(\mu K^2)}$  & $\rm{(\mu K^2)}$ \\ 
\hline
$70$ & $0.57$ & $1.13\times 10^{-3}$ & $1.08\times 10^{-3}$ & \\
\hline
$100$ & $0.41$ & $2.05\times 10^{-4}$ & $2.25\times 10^{-4}$ & \\
\hline
$143$ & $0.25$ & $4.52\times 10^{-5}$ & $4.86\times 10^{-5}$ &$1.2\pm 0.2 \times 10^{-5}$\\ 
\hline
$217$ & $0.16$ & $1.60\times 10^{-5}$ & $1.74\times 10^{-5}$ & $6\pm 1\times 10^{-5}$\\ 
\hline
$353$  & $0.33$ & $9.40\times 10^{-5}$ & $1.03 \times 10^{-4}$ &$1.9\pm 0.3 \times 10^{-3}$\\ 
\hline
\end{tabular}
\caption{\label{tab_PSP}
Poissonian contributions to the angular power spectrum for each frequency.
In the first column are reported the observational frequencies and in the second the detection threshold fluxes.
In the third column are reported the values of the Poissonian angular power spectrum for radio sources obtained 
 using the composite number count fits. In the fourth column are reported the values of the 
Poissonian angular power spectrum computed using the original De Zotti et al. model number counts instead of the composite ones.
In the last column are reported the values of the Poissonian term for infrared sources as predicted 
in \citep{PlanckCIB}.}
\end{table}

\subsection{Contribution from clustering term}

The Cosmic Infrared Background (CIB) is the integrated IR emission
over all redshifts of unresolved infrared star forming galaxies \citep{LagachePugetDoleCIB}. 
This contribution  is mainly that of starbust galaxies, Luminous InfraRed
Galaxies (LIRGs, $10^{11}L_\odot<L_{IR}<10^{12}L_\odot$) and Ultra
Luminous InfraRed Galaxies (ULIRGs, $L_{IR}>10^{12}L_\odot$), (with
small contribution from AGN)
\citep{LagachePugetDoleIR,LagachePugetDoleCIB,FernandezCondeI}. 
One of the complexity related to the CIB is that 
at different frequencies it accounts for contributions of IR sources from different redshift
ranges (e.g. \cite{FernandezCondeI}). 

The clustering power
spectrum is given by \cite{KnoxCooray2001,FernandezCondeI,Penin}: 
\ba
C_\ell^{\rm Clust}&=&\int
dz \frac{a^2(z)}{d_A^2}\frac{dr}{dz}\times\nonumber\\ 
&&\bar j^2_\lambda(z)P_g(k)|_{k=\ell/d_A}(z)\,,
\label{Clclust}
\ea 
where $r$ is the proper distance, $d_A$ is the comoving angular diameter distance, $k=\ell/d_A$ derives from the
Limber approximation, the $\frac{dz}{d_A^2}\frac{dr}{dz}a^2(z)$ term
takes into account all geometrical effects and depends on the
cosmological model adopted, $\bar j^2_\lambda(z)$ is the mean galaxy
emissivity per unit of comoving volume and
$P_g$ is the galaxy three-dimesional power spectrum .  The emissivity of infrared galaxies can be
written as \citep{Penin}: 
\be 
j_\nu(z)=\left(a \frac{d\chi}{dz} \right)^{-1} \int_L S(L_{IR})\frac{dN}{dzd(lnL_{IR})}d(ln{L_IR})
\ee
where $\frac{dN}{dzd(lnL_{IR})}$ is the luminosity function and $S$ is the flux.
The clustering term as shown by Eq. \ref{Clclust} depends on the
cosmological model through 
the geometrical factor, and through the galaxy power spectrum.  
Another complication is given by the indirect dependence of the emissitivity
on the assumptions on the galaxy Spectral Energy Distribution
\citep{KnoxCooray2001,FernandezCondeI}.
Finally, the bias, relating the
the emissivity fluctuations to the dark matter density fluctuations, also affects the
clustering term and is poorly known. In view of the the theoretical 
complexity of the
clustering term, we rather used the observational clustering term
extracted directly from {\it Planck} maps at 217 and 353 GHz by
\cite{PlanckCIB}. In order to minimally paramatrize this contribution, we have fitted
the {\it Planck} best fit models.
Since the Planck data are not available for the channels below 217 GHz, 
the values for the clustering term for 100 and 143 GHz channels
were derived using the halo occupation distribution best-fit 
parameters of the 217 GHz channels (Table 7 of \cite{PlanckCIB}), 
combined with the emissivities at 143 and 100 GHz computed using the 
\cite{Bethermin2010} model.

The fit to the clustering term is given for all the frequencies by the following expression: 
\be
\ell(\ell+1)\frac{C_\ell^{\rm clust}}{2\pi}(\nu)\propto \Big(\phi(\ell, \nu)+ \beta(\nu)
\frac{\log(\ell)^{\alpha(\nu)}}{(100+\ell)^{\gamma(\nu)}}\Big)\,, 
\label{cldata}
\ee
 where $\phi(\ell, \nu)$ is a fifth order polynomial 
term $A_0/\ell^2+A_1/\ell+A_2+A_3
\ell+A_4 \ell^2+ A_5 \ell^3+ A_6 \ell^4$ accounting for the 
steepening at high $\ell$, whereas the low $\ell$ shape is dominated by a complex logarithmic form.
The coefficients $A_{0-6}$ depend on the frequency.
The first step to obtain a fit to the clustered term was derive a spectral shape, namely
the combination of a polynomial component with the logarithm form.
Then we fitted the form given in Eq. \ref{cldata} for each frequency first fitting 
the logarithmic function and then the polynomial function. Finally, we have
derived the frequency dependence of each coefficient $A_{0-6}$ and each exponent by 
fitting to the different frequency. The resulting fit to the power spectrum is given by:
\ba
\ell(\ell+1)\frac{C_\ell^{\rm clust}}{2\pi}&=&\Sigma_{i=0,6} A_i \ell^{(i-2)}+A_7(\nu)
\frac{\log(\ell)^{A_8(\nu)}}{(100+\ell)^{A_9(\nu)}}\,,\nonumber\\
A_i&=&\Sigma_{j=1,3} a_i^j \nu^j\,,
\label{clfit}
\ea

where the first sum, on $i$, is on the coefficients of the polynomial and the
second, in $j$ is on the polynomial dependence on the frequency.
In Table 7 are reported the coefficients $a_i^j$.
Such complex formula insures an accuracy of the fit
at $< 1\%$ level in all $\ell$ range,
and at $<0.1\%$ in about $90\%$ of the considered $\ell$ range.

\begin{figure}
\includegraphics[scale=0.8]{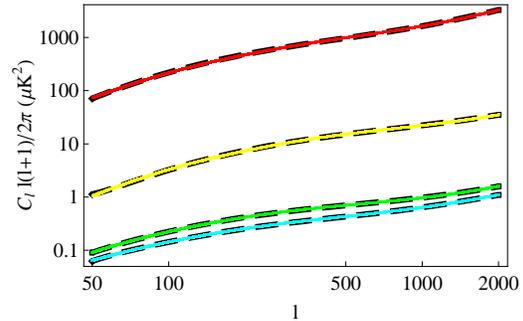}
\caption{Comparison between the best fit models from data (for 217 and 353) and extrapolations (for 100 and 143) 
of the infrared source clustering (dashed) and the fits (solid line) 
for the various frequency channels. Colors: Cyan
  $100$ GHz, Green $143$ GHz, Yellow  $217$ GHz, Red $353$ GHz.} 
\label{CIRBC}
\end{figure}
Finally, provided the expression Eq. \ref{clfit} giving the 
frequency-dependent CIB clustering power spectrum, the parametrization of the
clustering term can be simply written as:
\be
C_\ell^{\rm clust}(\nu)=A_{\rm CL}C_\ell^{\rm clust}(\nu)\nonumber\\
\ee
where the amplitude parameter $A_{\mathrm{CL}}$  accounts for deviations of 
the real signal from our parametrisation.

\section{Methodology}

\begin{figure}
\includegraphics[scale=0.9]{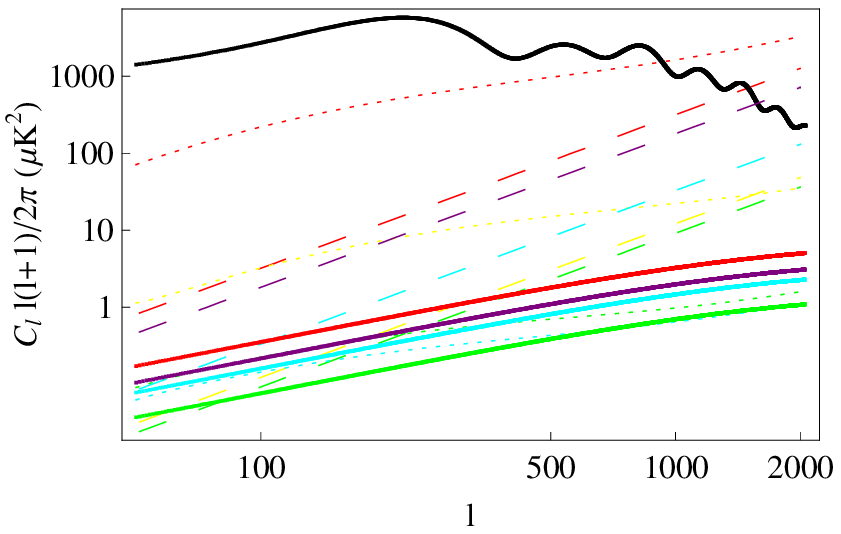}\\
\includegraphics[scale=0.9]{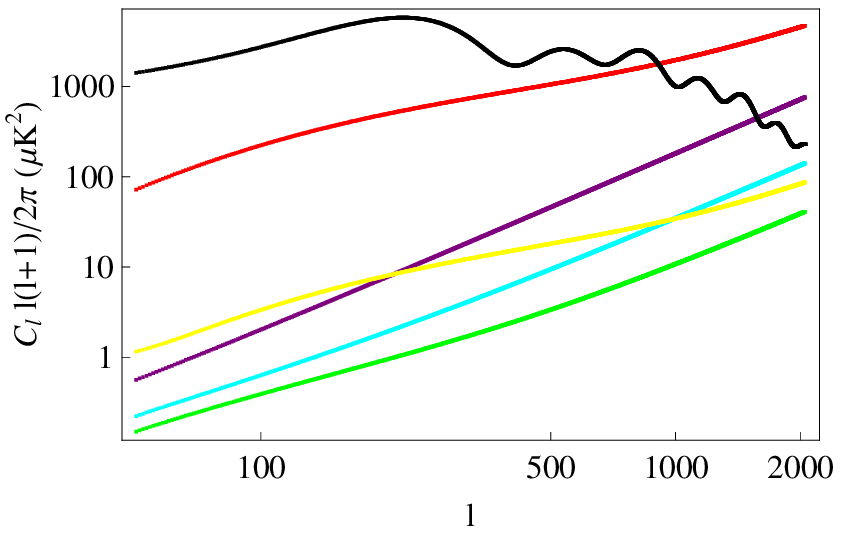}
\caption{Comparison of the extragalactic signals with primary CMB (black solid line) for the different frequencies. The line color represent the 
frequency: $70$ GHz (purple), $100$ GHz (cyan), $143$ GHz (green), $217$ GHz (yellow), $353$ GHz (red).
In the upper panel we show the comparison for each frequency of the three extragalactic 
signals: dotted is the  clustering term, dashed is the 
Poissonian term and solid is TSZ. In the lower panel we show the comparison of the total astrophysical contribution with the CMB for each frequency.
The solid lines are the sum of the TSZ, Poissonian and clustering contribution and follow the color scheme above.}
\label{ClvsFG}
\end{figure}

In the previous section we have detailed how we construct a frequency
dependent parametrization for the point sources and TSZ contributions to the 
angular power spectrum. 
In Fig. \ref{ClvsFG} we display each contribution
together with the primary CMB spectrum for the five
frequencies of interest. As anticipated, we note that the higher
frequency channels are dominated by the clustering term contribution,
the $353$ GHz channel being the most contaminated. The lower
frequencies are in turn dominated by the Poissonian term contribution.
We also note that at small scales the TSZ contribution remains
subdominant with respect to that of the point sources.  As seen in
Fig. \ref{ClvsFG}, displaying the comparison between the sum of all
contributions with the primary CMB, the $353$ GHz channel is the most
contaminated due to the clustering contribution, whereas the cleanest
channel is the $143$ GHz due to the low contribution of both
clustering and Poissonian terms.
We show in Table \ref{rms} the relative importance of the two main contributions
of clustering and Poissonian expresssed in terms of the root mean square:
\be
\sigma= \sqrt{\Sigma_l (2 l+1) b_l^2 C_l/(4\pi)}\,.
\ee
We note again how the high frequencies have a dominant clustering contribution whereas
for the 100 and 143 GHz the dominant contribution is the Poissonian term.

\begin{table}
\centering
\small\addtolength{\tabcolsep}{-4pt}
\begin{tabular}{|c|c|c|c|}
\hline
$\rm{Channels}$ & $\sigma^{\rm Clustering}$  & $\sigma^{\rm Pois,Radio}$  &$\sigma^{\rm Pois, IR}$ \\ 
\hline
 & $Jy/\sqrt{sr}$  & $Jy/\sqrt{sr}$  & $Jy/\sqrt{sr}$ \\ 
\hline
$100$ & $193.906$ & $789.309$ & \\
\hline
$143$ & $427.572$ & $786.607$ &$405.302$\\ 
\hline
$217$ & $2890.06$ & $873.219$ & $1690.98$\\ 
\hline
$353$  & $14927.1$ & $1312.22$ &$5899.56$\\ 
\hline
\end{tabular}
\caption{\label{rms}
Rms of the different contributions compared.
In the first column are reported the interested frequencies.
In the second column are reported the values of the clustering rms, in the third the Poissonian rms
 for radio sources and in the fourth the rms values for the Infrared Poissonian.}
\end{table}

Since we are interested in small scale anisotropies it is possible to 
avoid the component separation process to remove diffuse emission from our galaxy
and use instead single frequency angular power spectra derived by masking 
the Galaxy and the resolved point sources. 
The Galaxy may contribute even after masking  with high latitude emissions
and residuals which may remain depending on the aggressiveness of the mask used.
In this work, we neglect the possible residuals, after masking, from Galactic contamination.

We combine the angular power spectra of the microwave sky at the
considered frequency channels in a single power spectrum that we use
to estimate the cosmological parameters. 
An optimal method to combine channels is the inverse noise variance
weighting scheme \citep{EfstathiouGratton}.  This weighting scheme
indeed gives the larger weights to the channels with higher
resolutions and lower noise levels. This scheme is particularly
adapted to the case of high multipole analysis where the noise and the
beam smearing start to dominate. We therefore use a weighted linear
sum of single frequency power spectra: \be
C_\ell^\mathrm{tot}=\Sigma_i w_i(\ell) C_\ell^i\nonumber \ee where the
summation in $i$ is over the frequencies.
 
For simplicity, the noise is modelled as an isotropic Gaussian white noise with
variance $\sigma_N^2$ on the total integration time. 
We define the beam function as:
\be
b^2_\ell = e^{-\ell(\ell + 1)(0.425\, \mathrm{FWHM}/60\, \pi/180)^2};
\ee
where FWHM is the full width half maximum of the channel-beams in
arcminutes. We illustrate our method on a Planck-like case taking the beams and noise characteristics from \cite{HFIDPC},\cite{LFIFlight}. Moreover, the noise has been divided by $2$ to account for the approved extension of the mission. The numbers are summarised in Table \ref{PlanckPer}.

We can thus define the noise function as:
\be
N_\ell=\frac{(\sigma_N)^2\Omega_\mathrm{pix}}{b^2_\ell}\,,
\ee
where $\Omega_\mathrm{pix}$ is the area corresponding to the pixel. Note that 
the total noise contribution to the angular power spectrum  considers also the contribution of the observed sky fraction 
$\Delta C_l/Cl= N_\ell \sqrt{2/(f_{sky}(2\ell+1))}$. Assuming that the sky fraction in the different
channels is the same, the contribution of this term cancels in the weights.
With these definitions the weights for the single frequency spectrum
is given by: 
\be
w_i(\ell)=\frac{\frac{1}{N^2_{\ell_i}}}{\Sigma_j\frac{1}{N^2_{\ell_j}}}\,,
\ee
where $i$ stands for the specific channel whereas $j$ runs over all channels.

\begin{table}
\small\addtolength{\tabcolsep}{-4pt}
\begin{tabular}{|l|c|c|c|c|c|}
\hline
Instrument & \multicolumn{1}{|c|}{LFI} &\multicolumn{4}{|c|}{HFI} \\
\hline
Center frequency GHz & 70 & 100 & 143 & 217 & 353   \\\hline
Mean FWHM (arcmin)  &13 & 9.88 & 7.18 & 4.71 & 4.5 \\
\hline
$\Delta T$ per pixel ($I$) & 24.2525 & 8.175 & 5.995 & 13.08 & 54.5 \\
\hline
$\Delta T$ per pixel ($Q\, \& U$) & 34.6075 & 13.08 & 11.1725 & 24.525 & 103.55 \\
\hline
\end{tabular}
\caption{\label{PlanckPer}
Planck channel performance characteristics \citep{LFIFlight,HFIDPC} for the nominal mission.}
\end{table}
\begin{figure}
\begin{tabular}c
\includegraphics[scale=0.9]{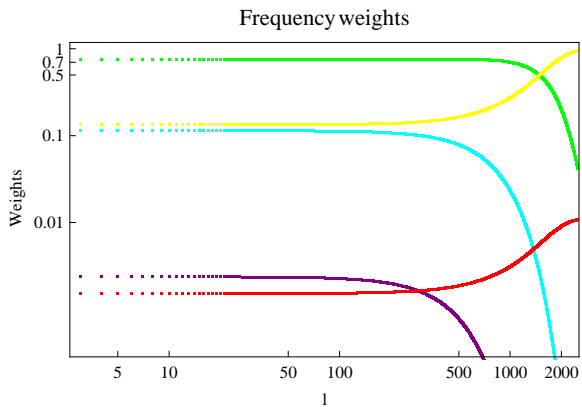}\\
\end{tabular}
\caption{Weights of the different frequencies with the inverse noise variance scheme: 
purple $70$ GHz, cyan $100$ GHz, green $143$ GHz, yellow $217$ GHz, red $353$ GHz.} 
\label{NUOVIINVERSE}
\end{figure}
We show the weights obtained using the inverse noise variance scheme
in Fig. \ref{NUOVIINVERSE}. We note 
how the dominant contribution among the various combinations is given by the
$143$GHz and $217$ GHz channels which have higher resolution and lower
noise level, whereas channels with lower angular resolution have very little weights
like the $70$ GHz whose contribution is strongly subdominant within this approach.

\section{Estimating the cosmological and extragalactic parameters} 

We now show the results of our analysis where we have applied, to the
 mock data, our approach of frequency-dependent minimal 
parametrisation of the extragalactic sources. 
We first show the capability of our approach to perfectly recover 
the input cosmological 
parameters in presence of extragalactic contamination and exhibit
the importance of the wide frequency range for the characterization 
of the extragalactic signals by comparing the results for different 
frequency combinations.
Then we show the impact of the astrophysical contributions 
to the power spectrum and to the cosmological parameters and their errors. 
We finally test the robustness of our approach by investigating the dependence of the results on the
astrophysical models we used. We will show also the results in the extended parameter space in particular we
considered the case of a $\Lambda$CDM with running of the spectral index and tensor modes.\\

We developed an extension of the publicly available CosmoMC code \citep{cosmomc}
which, in addition to estimating the cosmological parameters,
estimates at the same time the contributions of a
Poissonian term, a clustering term and a TSZ term.  This
is done by deriving a minimal number of amplitude terms, namely
$A_{\mathrm{SZ}}\,,A_{\mathrm{PS}}\,,A_{\mathrm{CL}}$, which
represent the departures from the input point source and TSZ models.    

In practice, we have modified the CAMB code \citep{camb}, version May 2010, 
to compute for each given frequency channel the sum of point-source and
TSZ spectra as provided by our parametrization (Sect. 2) and
the standard CMB angular power spectrum.  The resulting total power
spectra at the different frequencies are then combined in a single
effective power spectrum in the CosmoMC code.  

\begin{figure}
\includegraphics[scale=0.3]{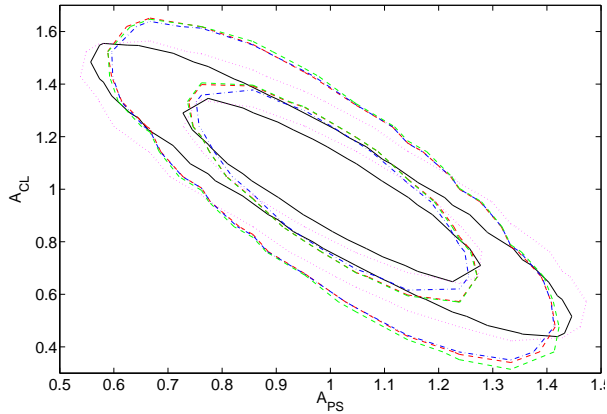}\\
\caption{Comparison of the results for the two main astrophysical parameters of the analysis obtained  with 
different frequencies, always combined with the inverse noise variance scheme: 
black (solid) is 70-353 GHz, red (short dashed) is 143-217 GHz, green (long dashed) is 100-217 GHz, blue (double dot dashed) is 70-217 GHz and the pink (dotted) line is the 100-353 
GHz. 
}
\label{FCF}
\end{figure}

We first explore the frequency combination and its effect on the
recovery of the cosmological and the astrophysical parameters. The latter 
are represented by the amplitudes\footnote{Note that our treatment of the TSZ contamination samples 
on all the physical factors multiplying the template $\hat C_\ell$ in Eq. \ref{SZR}, i.e. $A_{\mathrm{SZ}} \,, 
\sigma_8 \,, \Omega_b \,, h$. Therefore our approach differs from that adopted by the WMAP team which samples only on the overall 
amplitude multiplying the Komatsu-Seljak template \citep{KomatsuSeljak}.} $A_{\mathrm{SZ}}$, $A_{\mathrm{PS}}$ and 
$A_{\mathrm{CL}}$. We have
considered the results using five combinations of frequencies: 143 and 217 GHz,
100 to 217 GHz, 100 to 353 GHz, 70 to 217 GHz, and finally all the
frequencies under consideration 70 to 353 GHz, with a 
sky coverage of $f_\mathrm{sky}=0.85$. In order to establish
which frequency combination recovers the cosmological and astrophysical
parameters best, we have performed for each combination an MCMC analysis of mock
data using a $\Lambda$CDM-based CMB plus the astrophysical
contributions. For the latter we used the ``nominal'' contributions,
which means all the amplitude terms were set to one
($A_{\mathrm{SZ}}=A_{\mathrm{PS}}=A_{\mathrm{CL}}=1$). All the channel
combinations perfectly recover the cosmological parameters. The major
differences are in the recovery of the astrophysical parameters,
$A_{\mathrm{SZ}}$, $A_{\mathrm{PS}}$ and $A_{\mathrm{CL}}$, see 
Fig. \ref{FCF}. More specifically, the simplest
combination of the 143 and 217 GHz channels recovers the Poissonian
term with the smallest errors but it gives large errors for the
clustering term. By contrast, the combination of all considered
channels, 70 to 353 GHz, gives the largest errors for the
Poissonian term but significantly reduces the errors on the clustering
term.  The clustering term being the most uncertain with respect
to the Poissonian one, we chose to use the combination of the five channels
in order to improve the constraints on this major contribution
to the microwave sky at small angular scales.

\subsection{Standard $\Lambda$CDM model}

\begin{figure}
\begin{tabular}c
\includegraphics[{width=140mm,height=90mm}]{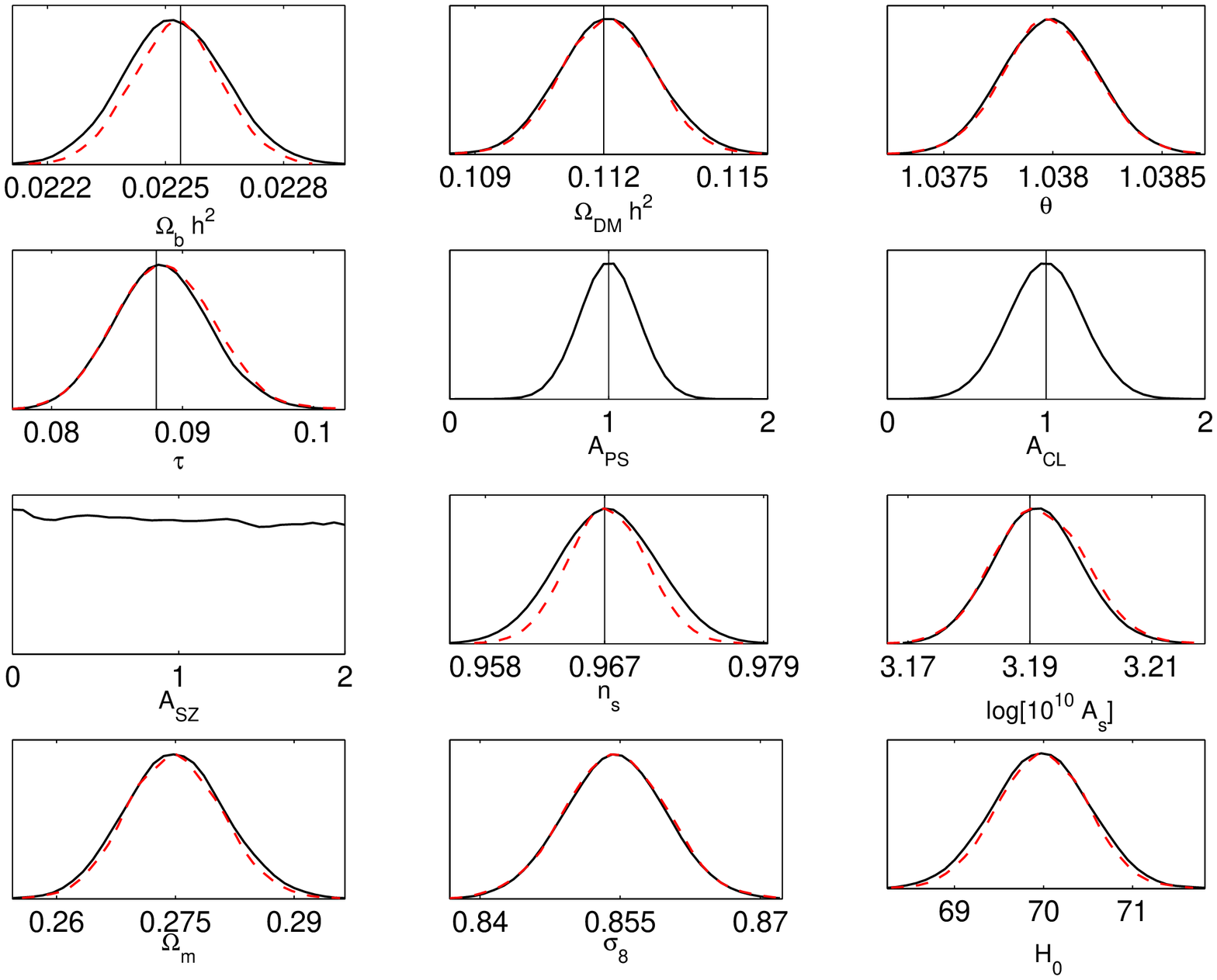}\\
\includegraphics[{width=180mm,height=140mm}]{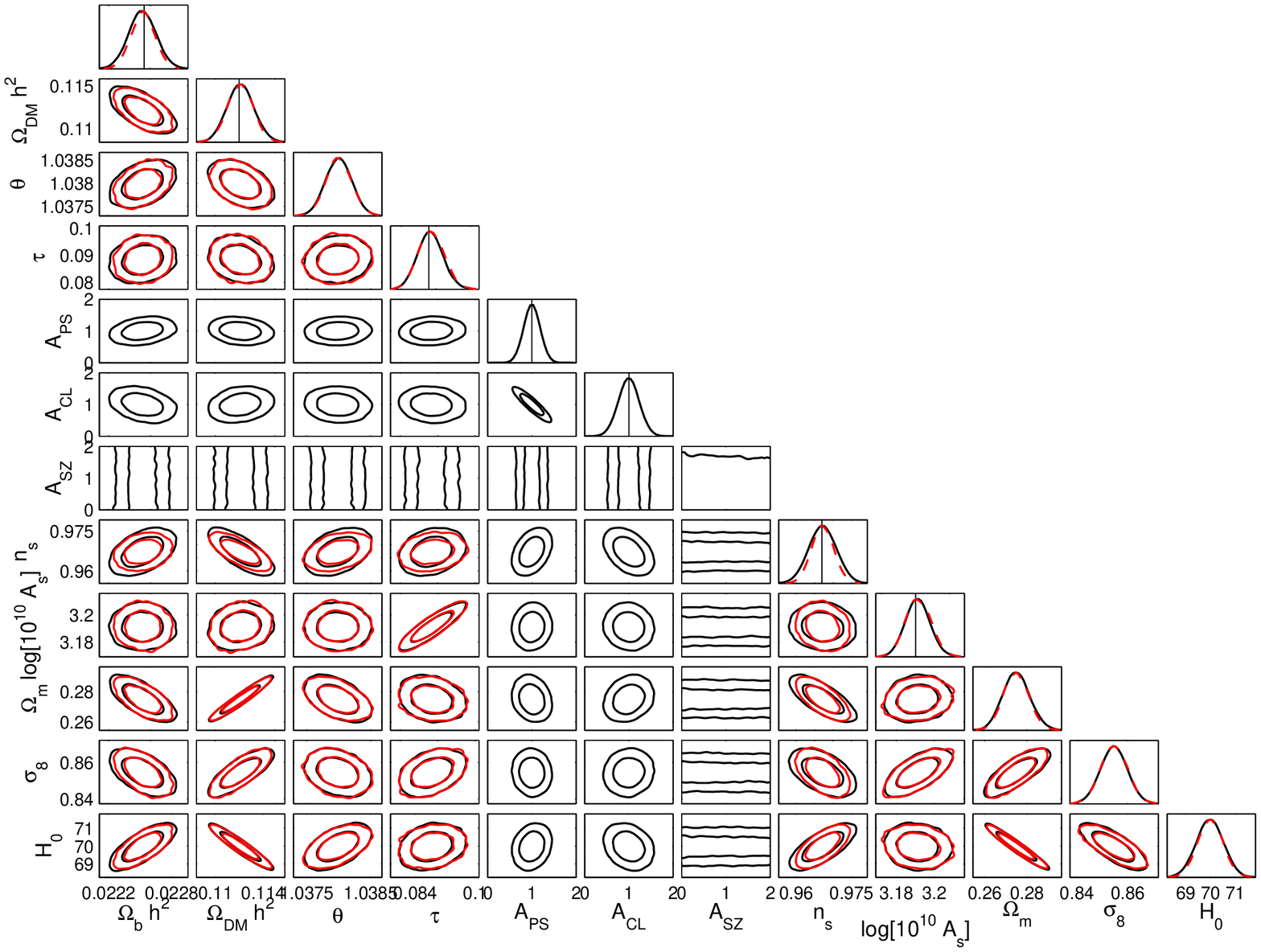}
\end{tabular}
\caption{Cosmological (and astrophysical) parameters with and without varying the astrophysical contribution parameters. 
The five frequency channels are combined with the inverse noise variance weighting scheme. 
The black (solid) curve represents the analysis which in addition to cosmological parameters varies also the astrophysical ones 
 whereas the red (dashed) curve represents the analysis with astrophysical parameters fixed to the fiducial model values.
In the lower panel we show also the triangle plot which is useful to evidence possible degenerations.
The curves are the $68\%$ and $95\%$ confidence level.}
\label{NOMINALDPC}
\end{figure}

We performed the analysis in the framework of a standard $\Lambda$CDM cosmological 
model. The input
mock $C_\ell$ were produced with nominal values for the point source and
TSZ contributions, namely $A_{SZ}=A_{PS}=A_{CL}=1$. Two cases are compared: Fixed values for the 
astrophysical parameters (fixed to one), and free values for $A_{\rm SZ}$,  
$A_{\rm PS}$, and $A_{\rm CL}$. 
The results are displayed in Fig. \ref{NOMINALDPC} and summarised in Table
4 (second column is the case with astrophysical parameters fixed to one, the third column is the case
with astrophysical parameters variation), together with the input parameters (fourth column) .
We show that the input
cosmological parameters are accurately recovered in both cases. 
When the astrophysical parameters $A_{\rm SZ}$,  
$A_{\rm PS}$, and $A_{\rm CL}$ together with the cosmological parameters are let free 
the errors bars are slightly enlarged. The astrophysical
parameters, $A_{\mathrm{PS}}$ and $A_{\mathrm{CL}}$, which
characterize the point sources (Poissonian and clustering terms) are
recovered well but they exhibit a strong degeneracy. 
This degeneracy is expected, in fact the Poissonian and the clustering
terms have different shapes but they both contribute additively to increase the measured power 
at high multipoles. The results are the same when we use the Tucci et al. model rather than out hybrid model for the radio contribution.
The amplitude of TSZ spectrum in turn remains unconstrained due to 
its subdominance with respect to the other two contributions of point sources. 
We will thus not discuss it further.

We now explore the robustness of the approach, in terms of parameter
recovery, with respect to the models we use for the point source
contributions. We do not vary the TSZ signal as it is subdominant and
cannot be recovered. For the effects of TSZ on the cosmological parameters in
absence of point sources, we refer the reader to \cite{TaburetBias,TaburetSZ}. 
We generated mock data including the
CMB, the nominal TSZ contribution and the point source contributions
(Poisson and clustering terms). The later have been chosen to differ
significantly from the nominal contributions.

\begin{figure}
\centering
\includegraphics[{width=85mm,height=100mm}]{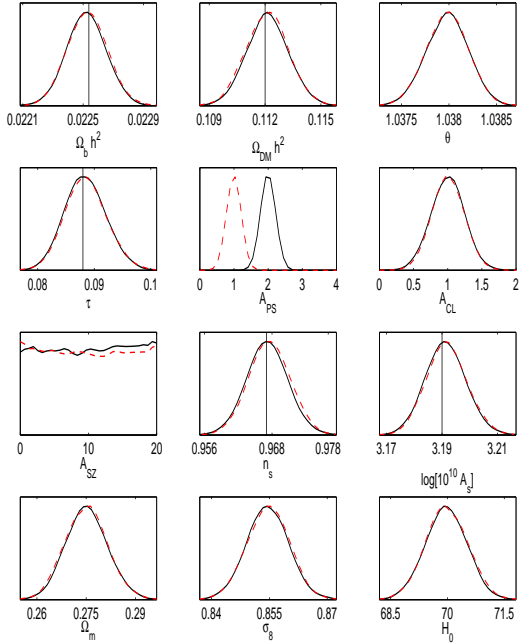}
\caption{Robustness test part 1. Cosmological and astrophysical parameters
with different astrophysical signals with respect to the fiducial model. 
The black (solid) curve are the results of the analysis which considers a Poissonian term which is two times its
fiducial value. The red (dashed) curve represent the analysis with a clustering term for the 100 and 143 GHz
doubled with respect to the fiducial value.}
\label{PSV}
\end{figure}

We first focus on the Poissonian term and multiplied the nominal
Poissonian term by two in the mock data for each frequency channel
leaving untouched the other contributions.  In Fig. \ref{PSV} we
present the obtained cosmological and astrophysical parameters.  It is
worth noting that even with a much larger Poissonian contribution the
cosmological and the astrophysical parameters, $A_{\mathrm{PS}}$ and
$A_{\mathrm{CL}}$, are perfectly recovered.
\begin{table}
\centering
\small\addtolength{\tabcolsep}{-4pt}
\begin{tabular}{|l|c|c|c|}
\hline
Parameter & Best Fit value I & Best fit value II & Input value \\
\hline
$\Omega_b h^2$&$0.02251\pm 0.0001$ & $0.02254\pm 0.0001$ & $0.0225$ \\
$\Omega_c h^2$&$0.1121\pm 0.001$&$0.1122\pm 0.001$ & $0.112$ \\
$\tau$&$0.0885\pm 0.004$ &$0.0874\pm 0.004$ & $0.088$ \\
$n_s$&$0.967\pm 0.003$&$0.966\pm 0.004$ & $0.967$ \\
$\log[A_s 10^{10}]$&$3.19\pm 0.007$ &$3.19\pm 0.007$ & $3.19$ \\
$H_0$&$69.96\pm 0.5$&$69.94\pm 0.54$ & $70$ \\
$A_{PS}$&$-$&$0.99\pm 0.17$ & $1$ \\
$A_{CL}$&$-$&$1.01\pm 0.22$ & $1$ \\
$A_{SZ}$&$-$&$-$ & $1$ \\
\hline
\end{tabular}
\label{bestfit} 
\caption{Comparison between the resulting best fit and the input
values for the cosmological and extragalactic parameters.
The fiducial model for the mock data is reported in the fourth column.
In the second column are given the parameters recovered  when taking into account the
  nominal contributions of the point sources and TSZ signals (note that the $A_{SZ}$ remains unconstrained). 
In the third column the parameters recovered  when varying also the parameters of the
 contributions of the point sources and TSZ signals.
The CMB plus extragalactic contribution power spectrum 
obtained using the best fit values and the one generated using the fiducial model values do
  not differ by more than 0.2\%. }
\end{table}
\onecolumn

\twocolumn
\par\bigskip
If we now focus on the clustering term, many tests with respect to the
nominal model can be envisaged to test the effect of the parametrisation on the parameter
recovery. We test the robustness of the results, or in other words the sensitivity to
the clustering term, with respect to the
frequency dependence of the clustering term by varying not only the
amplitude but also the frequency dependence of the clustering
term. The clustering term at 353 GHz being very well constrained by the
present {\it Planck} measurements of the CIB, we choose not to vary it. 
The CIB measurements at 217 GHz by {\it Planck}, in turn, have larger errors (see \cite{PlanckCIB}) and the values
adopted in the parametrisation for the 
100 and 143 GHz channels are empirical extrapolation based on the 217 GHz measure. 
Therefore it is very likely that at these frequencies the clustering term may be significantly
different from what we used in the model. For this reason, we chose to vary these three 
frequencies.

\begin{figure}
\includegraphics[{width=85mm,height=90mm}]{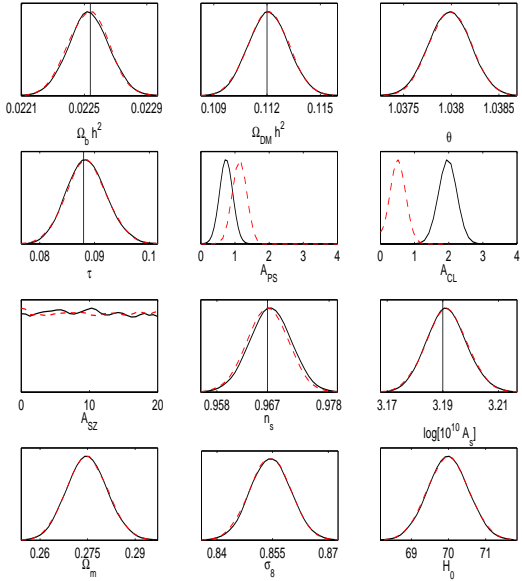}\\
\caption{Robustness test part 2. Cosmological and astrophysical parameters
with different astrophysical signals with respect to the fiducial model. 
The black (solid) curve are the results of the analysis which considers the 217 GHz clustering term double its
fiducial value. The red (dashed) curve represent the analysis with a clustering term for the 217 GHz half its fiducial value.}
\label{CL217}
\end{figure}

We first analyse a case where we multiply by two the
clustering contribution of the 100 GHz and 143 GHz channels. 
In Fig. \ref{PSV} the results show that this modification does not 
affect significantly the cosmological nor the astrophysical parameters. 
The reason is that at 100 and 143 GHz the clustering term
is not the dominant astrophysical component and therefore in the frequency combined
spectrum the variation of the clustering term for these two frequencies does
not modify the final result. In the inverse noise variance weighting 
combination scheme the major contribution to the clustering term comes from
the 217 GHz channel which has a strong clustering component and at
the same time is the dominant channel at high multipoles in the frequency combination.  
Since the 217 GHz measurements of the clustering still allow a rather significant variation, we 
decided to test the effects of a variation of the 217 clustering contribution by varying it by a factor 2. 
The obtained contribution differs by several sigmas
from the present constraints obtained by {\it Planck}.  We see from
Fig. \ref{CL217} that the cosmological parameters are perfectly
recovered either when we multiply or when we divide the 217 GHz clustering contribution by 
two. Unsurprisingly, the recovered mean values of $A_{\rm CL}$ are multiplied/divided by a factor near
 two following to the varied contribution. Due to the degeneracy between the Poissonian and clustering 
terms, the mean values of $A_{\rm PS}$ are also shifted with respect to the input value.

We now test the robustness of the cosmological parameter recovery with respect to the
frequency dependence that we proposed in our parametrisation of the clustering term. To do so, 
we multiply by two the clustering terms of the 100 and 
143 GHz, whereas the clustering term at 217 GHz is multiplied and then divided by
two. Here again
the cosmological parameters are perfectly recovered in both cases whereas 
the amplitudes of the Poisson and clustering terms vary. The results obtained for the mean values
of $A_{\rm PS}$ and $A_{\rm CL}$ are the same as those resulting from varying only the 217 GHz contribution.
This comes from the dominance of the 217 GHz channel weight, in the channel combination, which in turn
imply a dominance of the clustering term in the analysis.

\begin{figure}
\includegraphics[{width=85mm,height=90mm}]{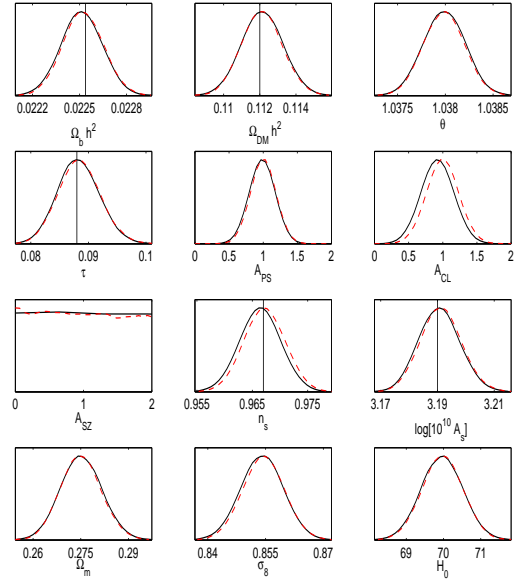}\\
\caption{Comparison of the cosmological and astrophysical parameters obtained with the analysis
which considers the standard De Zotti et al. based radio Poissonian (red curve) and
the one which considers the Tucci et al. based radio Poissonian in the mock data and the standard De Zotti et al.
 based Poissonian in the code (black curve).}
\label{TucciComp}
\end{figure}

We also tested the case where the Poissonian contribution is completely different from our model. 
To do so, we analyse the data using our hybrid model whereas we construct the mock data from the Tucci et al. model. 
By doing so, the Poissonian term varies from frequency to frequency whereas the IR contribution is the 
same as in the previous tests.
In Fig.\ref{TucciComp} we show that this different Poissonian contribution does not alter the cosmological parameters
 but the astrophysical ones are modified. 
In particular the Poissonian contribution moves along the degeneracy line and the clustering amplitude is shifted.


\subsection{Extended cosmological parameter space} 

We have shown the results of the cosmological and astrophysical
parameter recovery for a standard Lambda six-parameter cosmological
model. We now present some results obtained in the case of an
 extended cosmological parameter space.

\begin{figure}
\includegraphics[{width=85mm,height=90mm}]{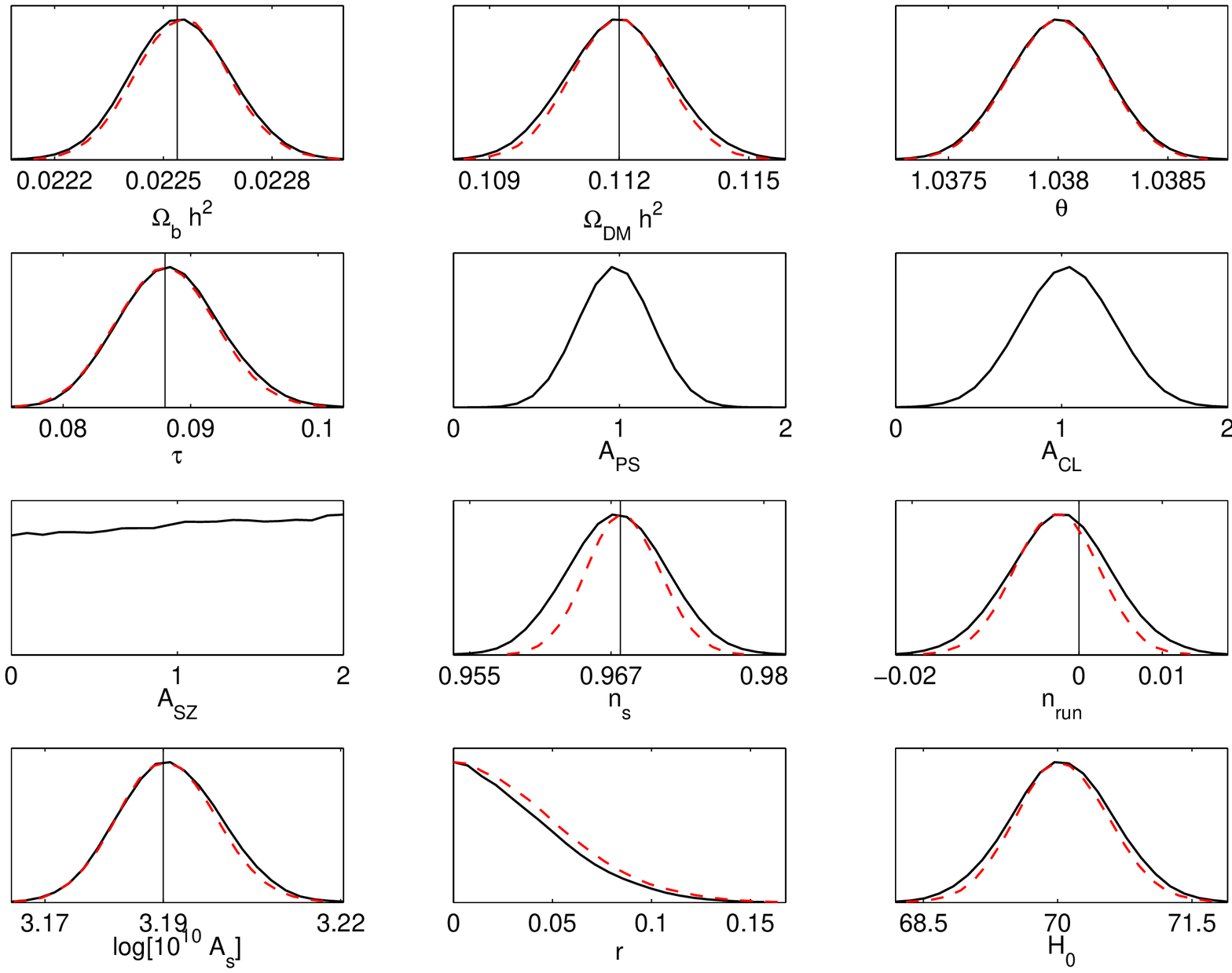}
\caption{Results of the analysis which considers together with cosmological and astrophysical parameters also the
 variation of the tensor to scalar ratio and the running of the spectral index.
Comparison between the results varying or not the astrophysical contributions:
the black (solid) curve considers the astrophysical signals whereas the red (dashed) curve represent the results with
 astrophysical contributions fixed to the fiducial value. Vertical bars are the input parameters.}
\label{GWRUNFG}
\end{figure}

We considered together with  the 
six standard parameters both tensor perturbations and the running of the spectral index,
varying the tensor-to-scalar ratio, $r$, and the running, $n_{\rm run}$, together with 
the six standard parameters and the three astrophysical ones. 
We have generated the mock data with both extra parameters, $n_{\rm run}$ and $r$ set to zero.  
We fix the tensor spectral index to the second order inflationary consistency condition 
$n_{\mathrm{T}} = -\frac{A_{\mathrm{s}}}{8}\left(2- \frac{A_{\mathrm{s}}}{8}-n_{\mathrm{s}}\right)$.
We included only the TE and EE mode in polarization we 
do not include the $B$-mode polarization power spectra in the exact form of the CMB likelihood we exploit, since 
the main limitation in that respect is given by the ability of removing diffuse foregrounds rather than instrumental noise. 
In Fig. \ref{GWRUNFG} we show the comparison of the
analyses which include the variation of the tensor-to-scalar ratio and the running of the
spectral index, with fixed amplitudes of the point source contributions (red, dashed, curve) 
and without fixing the amplitudes of the point source contributions 
(black, solid, curves). 

\begin{figure}
\includegraphics[scale=0.3]{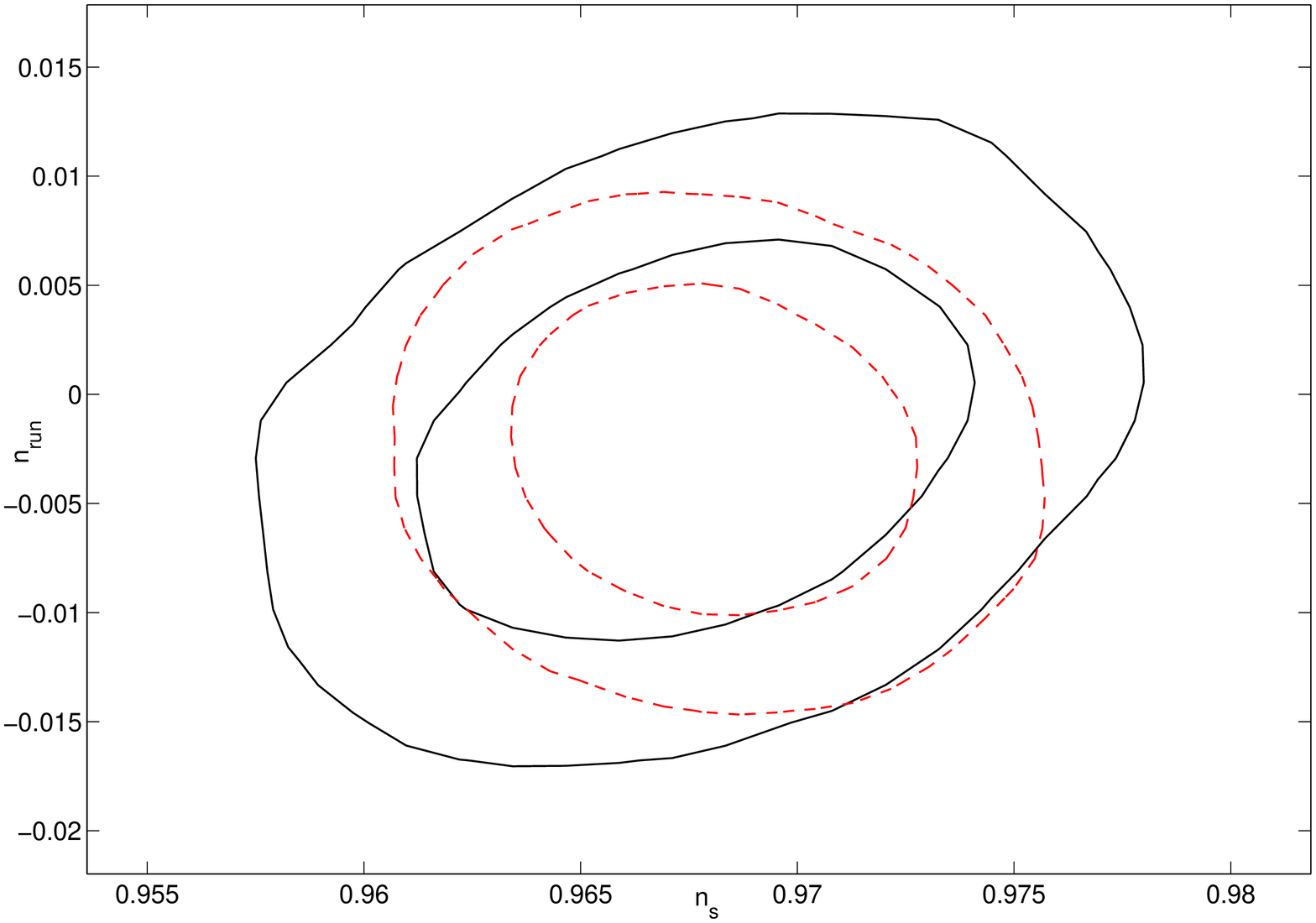}
\caption{Results of the analysis which considers together with cosmological and astrophysical parameters also the
 variation of the tensor to scalar ratio and the running of the spectral index.
Comparison between the results varying or not the astrophysical contributions:
the black (solid) curve considers the astrophysical signals whereas the red (dashed) curve represent the results with
astrophysical contributions fixed to the fiducial value.
We show the comparison of the two analysis for the running of the spectral index and the scalar spectral index.
Curves are the $68\%$ and $95\%$ confidence level.}
\label{GWRUNFG_2D}
\end{figure}

Similarly to the standard $\Lambda$CDM model, leaving the 
astrophysical contributions free enlarges the errors bars including those of the running of the 
spectral index. The bi-dimensional plot of the scalar spectral index versus its
running in Fig. \ref{GWRUNFG_2D} shows, in addition to the enlargement of the errors,
a degeneracy between the two cosmological parameters. 
The combined variation of the tensor-to-scalar ratio and  running, without the inclusion of $B$-mode polarization
shifts the mean of the running of the spectral index towards negative values with respect to input ones ($n_{run}=0$).

\begin{figure}
\includegraphics[{width=85mm,height=100mm}]{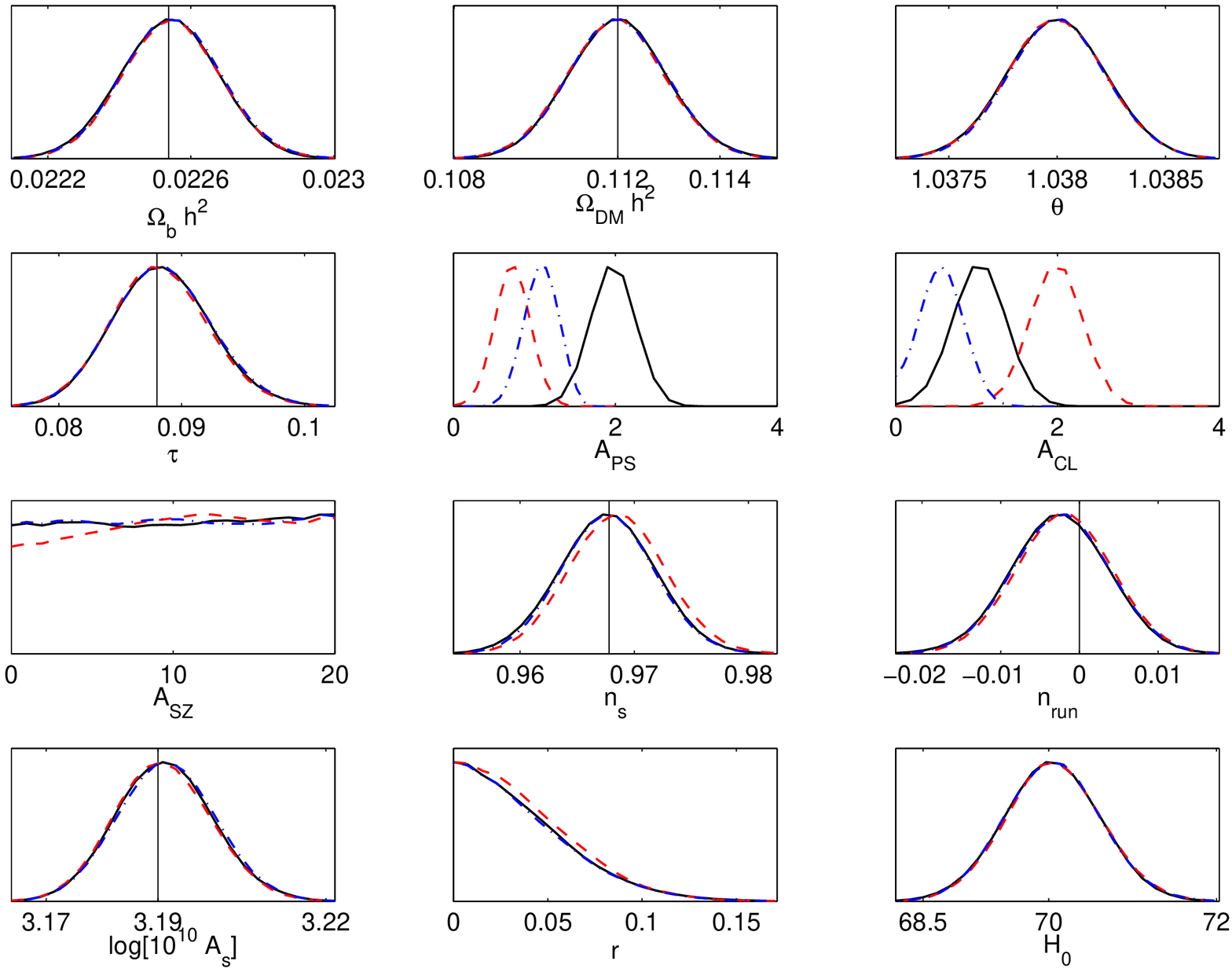}\\
\caption{Robustness test part 4. Cosmological and astrophysical parameters
with different astrophysical signals with respect to the fiducial model considering also the
tensor perturbations and the running of the spectral index. 
The black (solid) curve are the results of the analysis which considers a Poissonian contribution double its fiducial value.
The red (dashed) curve are the results of the analysis which considers the 100, 143 and 217 GHz clustering terms double their
fiducial value. 
The blue (dot-dashed) curve represent the analysis with a clustering term for the 217 GHz half its fiducial value whereas double
its fiducial value for 100 and 143 GHz.}
\label{FGV_RG}
\end{figure}

We repeated also the main robustness analysis in this extended parameter space. In particular we
repeated the three main cases: the case with the Poissonian contribution multiplied
by two, the variation of both amplitude and frequency dependence of the 
clustering term with the multiplication by two of the 100 and 143 and multiplying 
and dividing by two the 217 term.
In Fig.\ref{FGV_RG}  we report the results of the analysis.
We note how the recovery of the parameters is the same as in the standard astrophysical case 
showing how our method is robust also for this extended parameter space.

\section{Discussion and conclusions}

In the present study, we developed a multi-frequency approach of the analysis of the CMB
power spectrum aiming at obtaining unbiased cosmological parameters in the presence of 
unresolved extragalactic point sources and TSZ signal. 

We have considered the three main extra-galactic contributions to the CMB signal at
small scales, such as radio sources, infrared sources and SZ clusters. We have modelled 
these three signals in a minimal way using the observational constraints from {\it Planck}'s
early results \citep{PlanckRadio,PlanckCIB}. In particular, the radio and IR sources were 
modelled through a unique Poissonian term and a unique clustering term. 
In total, three 
amplitude terms suffice to account for and characterize the extra-galactic contributions. This minimal 
number of parameters was obtained by considering the frequency dependences within our 
parametrizations of the TSZ signal, and IR and radio point sources.

The use of a data-based physical model including the frequency dependence
of the astrophysical signals under consideration to build the parametrizations
of the small-scale contribution to the CMB is one of the original points of the
present work. Previous studies \citep{EfstathiouGratton,Millea} chose blind 
parametrizations of the signals which consider generic shapes and parametrize each frequency
with a different set of parameters.
Such an approach applied for a frequency channels, leads to 
either a rapid increase of the number of parameters necessary to 
describe the astrophysical signals, like in \cite{Millea}, or it requires the use of a limited
number of frequencies, like in \cite{EfstathiouGratton}. Other studies focused only on one single
contribution using more detailed and physical based parametrizations which involve
a higher number of parameters with respect to our approach (see for example \cite{SZGeorge} 
for the TSZ) or considered the frequency dependence on a much simpler level like for the clustering contribution
in \cite{IRCTemplate}. A different approach, relying directly on the data, has been used by the SPT 
team \citep{SPT2011,SPT}. They took advantage of the high resolution to acquire data on scales where 
the CMB is suppressed. They used the data on smallest scales to measure the amplitudes of their blind 
templates for the astrophysical signals and consequently to remove the associated contamination
from the CMB data.

We have applied our multi-frequency minimal parametrization of 
the extra-galactic contamination 
to estimate the cosmological parameters for 
a standard six-parameter $\Lambda$CDM model 
and for a model with tensor perturbations and a running of the 
spectral index, without including residuals 
from our Galaxy and without using the $B$-mode information. We 
show that the best 
combination of frequency is the one which considers the frequencies 
from 70 to 353 GHz. It allows to 
estimate the cosmological parameters and at the same to put the tightest 
constraints on the amplitude
of the IR clustering term. We show, both in the standard $\Lambda$CDM model 
and in the "extended-parameter" model,
that the input cosmological parameters together with the astrophysical 
ones (amplitudes of the TSZ, Poisson 
and clustering terms) are recovered as unbiased and with slighlty 
enlarged error bars. We show 
that the Poisson and clustering terms are degenerate. 

Our approach instead of blind parametrizations prefers to use the whole 
available information on point sources
and theoretical knowledge of the TSZ to have models of the 
astrophysical signals which are the most
similar as possible to the expected signal. 
The use of the frequency dependence and the technique used to 
fit the data
allowed a very restricted number of parameters to characterize 
the signals, in particular only a total of three parameters for all the signals. 

We have investigated for both cosmological models, standard and extended, the 
dependence of our approach and of the derived parameters on the accuracy of 
our model of extra-galactic contributions. In this way we tested its robustness 
against our theoretical priors on frequency dependence.
We tested different Poissonian and
clustering contributions by varying the associated terms both in amplitude and 
in frequency dependence. We showed that for both cosmological models
the cosmological parameters are recovered unbiased
even in situations where the mock data differ by several sigmas from the model.
The astrophysical parameters, amplitudes $A_{\rm PS}$ and $A_{\rm CL}$, are 
recovered well in the case of a ``real'' Poissonian term different from that of
the parametrisation. In the case of a different clustering term, the degeneracy 
with the Poissonian term induces a shift also in the Poissonian term.

We have provided a minimal parametrisation to take into account the extragalactic contaminations 
on small scales. Our minimal analysis can be extended 
to include additional contributions like residuals from the Galaxy contamination. However, it is ideal in the sense that systematic effects, beam uncertainties etc are not accounted for.

\begin{table}
\centering
\small\addtolength{\tabcolsep}{-4pt}
\begin{tabular}{|l|c|c|c|c|c|c|}
\hline
Channel & $70$ & $100$  & $143$  & $217$ & $353$\\
(GHz) &  &   &   &  & \\
\hline
$A^{\mathrm Low}$& 8.06 &7.76  & 7.47   & 7.16 & 6.81 \\
$B^{\mathrm Low}$& $10^{14}$ & $10^{14}$  & $10^{14}$   & $10^{14}$  & $ 10^{14}$ \\
$C^{\mathrm Low}$& 1 & 1  & 1   & 1 &1 \\
$\alpha^{\mathrm{Low}}$& 0.45 & 0.45 &   0.45   & 0.45 & 0.45 \\
$\beta^{\mathrm{Low}}$& 2.5 & 2.5 &  2.5   & 2.5 & 2.5 \\
$w_{\nu}^{\mathrm Low}$& 4.4 & 4.0 &  5.0   &  7.5 &  4.5 \\
\hline
$A^{\mathrm{Med}}$& 306.98 &224.99  & 241.08   & 203.69 & 165.85 \\
$B^{\mathrm{Med}}$& 20408 & 40000 & 111111  &   308642 &  501187 \\
$C^{\mathrm Med}$& 1 & 1  & 1   & 1 &1 \\
$\alpha^{\mathrm{Med}}$& 0.75 & 0.74 & 0.75   & 0.73 & 0.7 \\
$\beta^{\mathrm{Med}}$& 2. & 2. & 2.  &  2 & 1.9 \\
$w_{\nu}^{\mathrm Med}$&2.3 & 2.3 &  2.7   &  2.7 &  3.8 \\
\hline
$A^{\mathrm High}$& 58.41 & 22.66 & 70.89   & 23.12 &51.17 \\
$B^{\mathrm High}$& 1. & 0.91 &  3.52  &   1.48 &  0.93 \\
$C^{\mathrm High}$& 1 & 0.25  & 1   & 0.29 &1 \\
$\alpha^{\mathrm High}$& 0.75 & 0.85 & 0.8   & 0.85 & 0.75 \\
$\beta^{\mathrm High}$& 1.45 & 1.0 & 1.2   & 1.1 &  1.5 \\
$w_{\nu}^{\mathrm High}$&1.0 & 1.0 & 1.04   &  1.0 & 1.03 \\
\hline
\end{tabular}
\label{FPOI} 
\caption{The exponents $\alpha$, $\beta$ and the coefficients $A$, $B$, $C$ and $w_{\nu}$ for each term of the fits of the
 radio galaxies composite number counts. }
\end{table}

\begin{table}
\centering
\small\addtolength{\tabcolsep}{-4pt}
\begin{tabular}{|l|c|c|c|c|c|c|}
\hline
Channel & $70$ & $100$  & $143$  & $217$ & $353$\\
(GHz) &  &   &   &  & \\
\hline
$P_1$& $1.1 10^{-4}$ & $3.6\times 10^{-5}$ & $1.1\times 10^{-5}$ & $4.3\times 10^{-6}$ & $1.8\times 10^{-5}$ \\
$\chi_1$& 0.95  & 0.95  &  0.95  & 0.95  &0.95 \\
$a_1$& 0.34 &  0.34 &  0.34  &  & \\
$b_1$& 1 & 1 & 1     & 1 & 0.34 \\
$c_1$& 1.334 & 1.34 & 1.34    &  1.34 & 1\\
$d_1$&$-10^{14}$ &$ -10^{14}$ & $ -10^{14}$  & $-10^{14}$  & $ -10^{14}$ \\
$e_1$& 2.5 & 2.8 &  2.8   & 2.8 &2.8  \\
\hline
$P_2$&$ 5.9\times 10^{-3}$ &  $1.4 \times 10^{-3}$ & $ 3.6 \times 10^{-4}$ & $1.5 \times 10^{-4}$ & $4.1 \times 10^{-4}$\\
$\chi_2$&1.28  & 1.24 & 1.2  & 1.15   &1.2 \\
$a_2$& 0.64 & 0.62  & 0.6   &0.58  & 0.63\\
$b_2$& 1 & 1. &   1.  &  1. & 1.\\
$c_2$& 1.64 &  1.62 &  1.6   & 1.58 &1.63 \\
$d_2$& -20408.2 & -40000  & -111111.    &-308642. & -501187  \\
$e_2$& 2 & 2 &  2   &2 &  1.9 \\
\hline
$P_3$&$2.6\times 10^{-3}$  &$1.2\times 10^{-3}$ & $3.6 \times10^{-4}$ & $2.5 \times 10^{-4}$  & $4.5 \times 10^{-4}$ \\
$\chi_3$&1.25 & 1.35  & 1.3   & 1.35 &1.25 \\
$a_3$& 0.86 & 1.  & 1   & 1. & 0.83\\
$b_3$& 1. & 1.35 &  1.08   & 1.22727 & 1.\\
$c_3$&1.86  & 2.35 & 2.08    & 2.23 & 1.83\\
$d_3$&  -1 &  -3.64 &  -3.52   &  -5.10 & -0.93\\
$e_3$& 1.45 & 1 & 1.2    & 1.1 & 1.47\\
\hline
\end{tabular}
\label{ClPTerms} 
\caption{ The coefficients and exponents of the radio galaxy Poissonian contribution to the angular power spectrum Eq.\ref{RadioP}. }
\end{table}

\begin{table}
\centering
\small\addtolength{\tabcolsep}{-5pt}
\begin{tabular}{|l|c|c|c|c|c|}
\hline
 $a_i^j$&j & 0 & 1 & 2  & 3     \\
\hline 
 i    &  &     &   &    &     \\
\hline
 0    &  & 14541.6 & -298.31 & 1.84   &$ - 3.26\times 10^{-3}$ \\   
\hline
 1    &  &   -718.25     &15.14  &  $- 9.8\times 10^{-2}$  & $1.9\times 10^{-4}$ \\   
\hline
 2    &  &     19.52   & - 0.42  & $2.8\times 10^{-3}$   &$- 5.99 \times10^-6$  \\ 
\hline
 3    &  &    -0.19    & $4.2\times 10^{-3}$  & $ - 2.8\times 10^{-5}$  & $6.1\times 10^{-8} $ \\ 
\hline
 4    &  &  $ -2.\times 10^{-4} $    & $4.4\times 10^{-6}$  &  $- 3.03\times 10^{-8}$  &   $6.72\times 10^{-11}$ \\  
\hline
 5    &  &  $ 1.83\times 10^{-8}$    & $- 3.99\times 10^{-10} $ & $ 2.78\times 10^{-12}$  & $6.18\times 10^{-15}$ \\
\hline
 6   &   &   $ 4.58\times 10^{-13} $  & $- 9.72\times 10^{-15}$  &   $6.50\times 10^{-17}$ & $-1.36\times 10^{-19}$ \\
\hline
 7   &   &    -108.25   & 2.40  & $ - 1.67\times 10^{-2}$  & $ 3.66\times 10^{-5}$\\
\hline
 8   &   &      3.81 & $ - 6.03\times 10^{-2}$ & $ 3.87\times 10^{-4} $  & $-6.53\times 10^{-7}$ \\
\hline
 9   &   &    2.76   & $- 2.08\times 10^{-3}$  & $- 1.13\times 10^{-5}  $ & $ 4.07\times 10^{-8}$ \\
\hline
\end{tabular}
\label{ClusteringTable} 
\caption{ The coefficients and exponents of the fits for the clustering term Eq.\ref{clfit}. }
\end{table}

 Acknowledgements: We wish to thank Nicolas Taburet for useful discussions and for providing the TSZ power spectrum template
and Marco Tucci for provinding the number counts of 
\cite{Tucci}.This work is partially supported by ASI contract Planck-LFI
activity of Phase E2, by LDAP. NA wishes to thank IASF Bologna for hosting and DP thanks IAS for hosting. The simulations for this work have been carried on IASF Bologna cluster.
We wish thank the financial support by INFN IS PD51 for NA visit in Bologna.

\end{document}